\documentclass[11pt,a4paper]{article}

\usepackage[T1]{fontenc}
\usepackage[utf8]{inputenc}
\usepackage{lmodern}
\usepackage[a4paper,margin=2.5cm]{geometry}
\usepackage{graphicx}
\usepackage{amsmath}
\usepackage{amssymb}
\usepackage{float}
\usepackage{placeins}
\usepackage{cite}
\usepackage[hidelinks]{hyperref}
\usepackage{orcidlink}

\begin{document}

\begin{titlepage}

\centering

{\LARGE \textbf{OpenMRF: A Modular, Vendor-Neutral Open-Source Framework for Reproducible Magnetic Resonance Fingerprinting using Pulseq}\par}

\vspace{1.5cm}

{\large
Tom Griesler$^{1,2}$, Jannik Stebani$^{3}$, Sydney Kaplan$^{1,2}$, Ivaylo Angelov$^{3}$, Petra Albert$^{3,4}$, Martin Blaimer$^{3}$, Tobias Wech$^{5}$, Xiang Wang$^{6}$, Qingping Chen$^{6}$, Maxim Zaitsev$^{6}$, Zhibo Zhu$^{7}$, Qi Liu$^{7}$, Peter Martin$^{7}$, Jon-Fredrik Nielsen$^{1,2}$, Jesse I Hamilton$^{1,2}$, Peter Nordbeck$^{4}$, Nicole Seiberlich$^{1,2}$, Maximilian Gram$^{1,3,4,*}$
\par}

\vspace{1.2cm}

\begin{flushleft}
$^{1}$ Department of Radiology, University of Michigan, Ann Arbor, Michigan, USA\\
$^{2}$ Department of Biomedical Engineering, University of Michigan, Ann Arbor, Michigan, USA\\
$^{3}$ Experimental Physics 5, University of W\"urzburg, W\"urzburg, Germany\\
$^{4}$ Department of Internal Medicine I, University Hospital W\"urzburg, W\"urzburg, Germany\\
$^{5}$ Department of Diagnostic and Interventional Radiology, University Hospital W\"urzburg, W\"urzburg, Germany\\
$^{6}$ Division of Medical Physics, Department of Diagnostic and Interventional Radiology, University Medical Center Freiburg, Faculty of Medicine, University of Freiburg, Freiburg, Germany\\
$^{7}$ United Imaging Healthcare North America, Houston, Texas, United States of America\\[0.5em]
$^{*}$ Corresponding author
\end{flushleft}
\,\\
\,\\

\begin{flushleft}
\textbf{Author's ORCIDs:}\\
Tom Griesler: \orcidlink{0009-0003-3777-5483} \href{https://orcid.org/0009-0003-3777-5483}{0009-0003-3777-5483}\\
Maximilian Gram: \orcidlink{0000-0003-2184-3325} \href{https://orcid.org/0000-0003-2184-3325}{0000-0003-2184-3325}
\end{flushleft}

\,\\
\,\\

\begin{flushleft}
\textbf{Address for correspondence:}\\
Maximilian Gram, University Hospital W\"urzburg, Department of Internal Medicine I, Oberd\"urrbacher Stra\ss e 6, DE-97080, W\"urzburg, maximilian.gram@uni-wuerzburg.de
\end{flushleft}

\,\\
\,\\
\begin{center}
\Large
\textbf{Submitted to \textit{Magnetic Resonance in Medicine}}
\end{center}

\end{titlepage}

\noindent\rule{\textwidth}{0.5pt}
\begin{abstract}
\normalsize
\,\\
\textbf{Purpose:} Widespread adoption and methodological advancement of Magnetic Resonance Fingerprinting (MRF) are limited by the lack of unified, reproducible implementation frameworks and fragmented open-source tools. To address these barriers, we introduce OpenMRF - a comprehensive Pulseq-based solution - designed to enable consistent, reproducible, and transferable MRF research across vendors, sites, and field strengths.

\,\\
\textbf{Methods:} OpenMRF integrates modular Pulseq-based sequence design, Bloch-simulation-based dictionary creation directly from .seq files, and iterative low-rank subspace reconstruction. The framework was evaluated through digital phantom simulations, a multi-site ISMRM/NIST phantom study on Siemens MRI systems at 0.55\,T, 1.5\,T, and 3\,T as well as GE and United Imaging 3\,T platforms, and representative \textit{in vivo} acquisitions in the liver (0.55\,T), myocardium (1.5\,T), and brain (3\,T).

\,\\
\textbf{Results:} Simulations demonstrated high mapping accuracy in an ISMRM/NIST-like digital phantom, with low-rank reconstruction yielding deviations of $0.03 \pm 0.32$ \% ($T_1$) and $0.12 \pm 1.94$ \% ($T_2$). The multi-site phantom study yielded relaxation times consistent with reference values at all field strengths, with mean deviations of $-0.1 \pm 2.9$ \% ($T_1$), $-1.5 \pm 8.7$ \% ($T_2$), and $-4.0 \pm 7.2$ \% ($T_{1\rho}$). \textit{In vivo} acquisitions produced high-quality parameter maps across platforms and field strengths.

\,\\
\textbf{Conclusion:} OpenMRF provides a robust, open-source, end-to-end Pulseq-based solution for MRF that enables reproducible sequence implementation, physics-accurate dictionary simulation, and advanced reconstruction across vendors and field strengths. By providing a unified platform for method development, comparison, and multi-site validation, OpenMRF aims to accelerate reproducible and harmonized quantitative MRI research within the community.
\,\\
\end{abstract}
\noindent\rule{\textwidth}{0.5pt}

\,\\

\noindent\textbf{Keywords:} MR Fingerprinting, Open-Source, Pulseq, quantitative MRI, $T_{1\rho}$, cross-vendor
\newpage

\section{Introduction}

Magnetic resonance imaging (MRI) is a powerful and versatile modality routinely used in both clinical practice and biomedical research. Its ability to probe a broad range of tissue properties arises from the high degree of freedom in MRI sequence design. The interplay between gradient and radiofrequency (RF) channels enables a near-limitless variety of contrast mechanisms and spatial encoding strategies. Beyond conventional contrast-weighted imaging, quantitative MRI (qMRI) has emerged as a promising approach for extracting biophysical tissue parameters such as relaxation times ($T_1$, $T_2$, $T_2$*, or $T_{1\rho}$), diffusion coefficients, perfusion metrics, and chemical exchange properties with high spatial resolution \cite{Gulani2020, Hockings2020}.

With the precision offered by modern MRI hardware, research groups continue to drive rapid innovation across multiple vendor platforms. However, this progress also highlights the need for standardization and reproducibility in qMRI. For clinical translation, cross-site and cross-vendor comparability of quantitative parameters is essential. Initiatives driven by ISMRM, SCMR and QIBA have therefore established consensus protocols and guidelines for robust parametric mapping \cite{Keenan2018, Messroghli2018, Shukla-Dave2019}.

A major milestone in the evolution of qMRI was the introduction of Magnetic Resonance Fingerprinting (MRF) by Ma et al. \cite{Ma2013}, which enables rapid, simultaneous quantification of multiple tissue properties through transient signal encoding and dictionary-based pattern matching. Since then, MRF has been adapted to a wide range of anatomical targets and clinical applications \cite{Gaur2023}.

Parallel to the development of MRF, vendor-neutral, open-source pulse sequence frameworks have emerged. Platforms like Pulseq make it possible to generate human-readable, fully programmable sequence definitions in Matlab \cite{Layton2017} or Python \cite{Ravi2019}, with growing support for scanners from Siemens, GE, Philips and United Imaging \cite{Nielsen2018, Roos2025, GramISMRM2026b}. Pulseq has been applied in diverse qMRI studies, including $T_1$ mapping, diffusion imaging, and CEST experiments \cite{Gaspar2023, Liu2024, Herz2021}.

Multiple different simulation tools can be used for signal prediction or full emulation of acquisition pipelines based directly on Pulseq definitions. However, MRF poses specific challenges that are not fully addressed by existing simulation tools. Accurate modeling of slice profile effects, magnetization decay during long adiabatic pulses and specific conditions during spin-lock preparations is crucial for reliable and generalized dictionary generation \cite{Ma2017}. Existing frameworks such as MRzero \cite{Loktyushin2021}, JEMRIS \cite{Stoecker2010}, or KomaMRI \cite{Castillo-Passi2023} are powerful but not tailored to the specific requirements of MRF. A recent open-source cMRF sequence based on PyPulseq demonstrated successful transfer across Siemens platforms \cite{Schuenke2025}, but the accompanying dictionary simulation tools were tailored to a specific sequence design, and image reconstruction relied on sliding-window methods and lacked state-of-the-art reconstruction strategies.

To the best of our knowledge, no open-source Pulseq-based framework currently unifies flexible MRF sequence development, physics-accurate dictionary simulation, and advanced reconstruction in a single reproducible workflow. To address this gap, we present OpenMRF, a modular and extensible framework for Magnetic Resonance Fingerprinting built using the Pulseq standard. Developed collaboratively, OpenMRF provides end-to-end tools for sequence implementation, automated Bloch-based dictionary generation, and state-of-the-art reconstruction, enabling reproducible MRF studies across vendors and field strengths. To demonstrate its flexibility, we apply OpenMRF to three representative MRF protocols for brain, abdominal, and cardiac imaging and validate the framework across Siemens, GE and United Imaging systems at 0.55\,T, 1.5\,T, and 3\,T.

\begin{figure}[p]
    \centering
    \includegraphics[width=\textwidth]{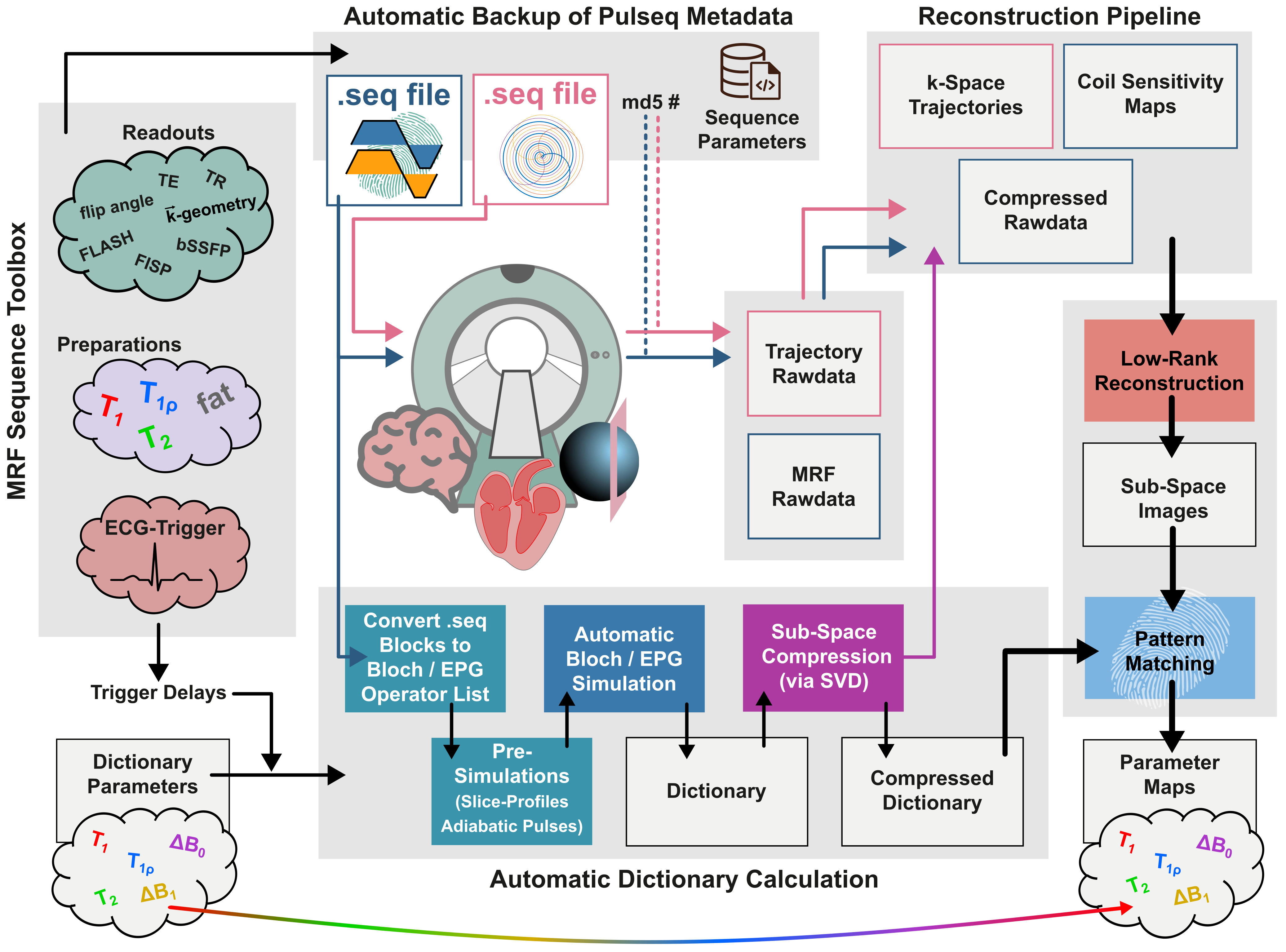}
    \caption{Schematic illustration of the OpenMRF framework for developing, simulating, and reconstructing Magnetic Resonance Fingerprinting (MRF) experiments based on the Pulseq standard. Pulseq .seq files define the sequence, including contrast preparation modules (e.g., inversion, $T_2$ preparation, spin-lock) and k-space trajectories. Optionally, a separate sequence can be generated for trajectory calibration. For dictionary generation, the .seq file is parsed into a compact instruction set and simulated with a Bloch-isochromat simulator that accounts for slice-profile effects, relaxation during adiabatic pulses, and spin-lock conditions. By compressing both the simulated dictionary and raw data onto a low-rank temporal subspace, the framework enables iterative subspace reconstruction, followed by dictionary matching to obtain quantitative parameter maps (e.g., $T_1$, $T_2$, $T_{1\rho}$, B0, $B_1^+$). All modules are linked through a standardized metadata concept that uses the Pulseq MD5 hash to connect raw data and sequence backups.}
    \label{fig:framework}
\end{figure}

\newpage
\section{Methods}

\subsection*{I. Features of OpenMRF}

The OpenMRF project aims to provide a vendor-neutral platform for developing, simulating, and reconstructing MRF sequences (\hyperref[fig:framework]{Figure 1}). Rather than focusing on a single clinical application, the framework is designed to support a broad range of use cases through modular tools that address the core requirements outlined in the Introduction. The entire source code is openly available on GitHub. OpenMRF builds upon the Matlab implementation of Pulseq and features modules for contrast preparation, advanced readouts, automated dictionary simulation, and model-based reconstruction. The following subsections describe the main components of the framework, while detailed documentation is provided on the accompanying documentation website (openmrf.org).

\subsubsection*{Readouts and trajectory calibration}
Although MRF is not limited to spiral readouts, spiral-out trajectories remain the most common choice due to their efficient k-space coverage \cite{Gaur2023}. In OpenMRF, the default readout option is a variable-density spiral \cite{Lee2003}. Users can select different projection-angle schemes, such as Equal-2pi or Golden Angle, though an approximated Golden Angle ordering with reduced angular increments is used by default. The readout can be operated in 2D slice-selective, 3D slab-selective (e.g. stack-of-spirals) or 3D mode, and spoiling can be configured as FLASH (RF and gradient spoiling), FISP (gradient-only spoiling), or bSSFP (no spoiling). Readout gradients are automatically equipped with rewinders, and gradient spoiling is performed along the slice-select direction.

A key feature of OpenMRF is the option for integrated trajectory calibration using the method of Robison et al. \cite{Robison2019}. This approach acquires each spiral interleave in an off-center slice using regular and inverted gradient polarity, enabling compensation of trajectory deviations caused by eddy currents and concomitant fields. For practical use, OpenMRF generates a dedicated Pulseq file for the calibration scans, allowing accurate measurement of the trajectories in a spherical phantom, which can then replace the nominal trajectory for more accurate image reconstruction.

\subsubsection*{Contrast preparation modules}
The original MRF approach relied on variable flip angles and repetition times to encode $T_1$ and $T_2$ \cite{Ma2013}, but modern MRF protocols typically use explicit preparation modules to enhance contrast encoding \cite{Gaur2023}. OpenMRF provides a modular library of state-of-the-art preparations that can be flexibly combined with any readout.

For $T_1$ encoding, adiabatic inversion pulses are implemented using hyperbolic secant designs generated with SigPy \cite{Ong2019}. $T_2$ preparation follows the structure used in several cardiac MRF sequences \cite{Hamilton2018}, employing BIR-4 pulses for 90-degree rotations and composite refocusing pulses. OpenMRF also supports spin-lock preparations for $T_{1\rho}$ encoding \cite{Wyatt2020, Sharafi2021, Velasco2022, Kaplan2025}. Both adiabatic half-passage pulses and B0/B1-robust techniques, such as composite and balanced spin-locking, are available \cite{Schuenke2017a, Witschey2007, Gram2021, Wan2025}. Additional modules include optional fat suppression and magnetization resets.

\subsubsection*{Automated Bloch simulation for dictionary generation}
Accurate signal simulation is central to MRF, and OpenMRF provides a fully automated workflow that derives all simulation parameters directly from Pulseq (.seq) files. The simulator converts each sequence block into a compact instruction set without any hard-coded parameters, ensuring reproducibility and generality.

In contrast to existing simulation tools \cite{Loktyushin2021, Stoecker2010, Castillo-Passi2023}, OpenMRF explicitly models physical effects that are known to impact MRF accuracy, including slice-profile imperfections and relaxation during long adiabatic pulses \cite{Hamilton2018}. Spin-lock conditions are automatically detected and the simulator switches to a modified Bloch formulation that incorporates $T_{1\rho}$ and $T_{2\rho}$ relaxation as well as B0 and $B_1^+$ deviations \cite{Mangia2009, Gram2024}.

The core simulator is based on the Bloch isochromat approach. For each sequence, a set of isochromats along the slice direction (typically 1000) is simulated, with two execution modes:
\begin{itemize}
\item Brute-force Bloch simulation: All RF pulses are simulated step-by-step at the hardware raster time, and slice profiles are explicitly modeled.
\item Accelerated Bloch simulation: RF pulses are approximated as instantaneous rotations according to pre-simulated slice profiles. Relaxation during adiabatic pulses is represented by precomputed transition matrices.
\end{itemize}

To improve computational efficiency, long delays are collapsed into single forward steps with adaptive time increments, and gradient moments are aggregated wherever possible. The simulator supports a wide range of tissue and hardware parameters, including: $T_1$ (longitudinal relaxation), $T_2$ (transverse relaxation), $T_2$’ (reversible transverse relaxation), $T_{1\rho}$/$T_{2\rho}$/dispersion (rotating frame relaxation), ADC (apparent diffusion coefficient), $\Delta B0$ and $\Delta B1^+$ (field deviations).

\subsubsection*{Low-rank reconstruction and pattern matching}
Quantitative parameter maps in OpenMRF are obtained by matching reconstructed signal evolutions to a dictionary generated from the corresponding .seq file. Two reconstruction strategies were used in this work and are available in the framework.

In the first approach, the dictionary $D \in \mathbb{C}^{N_R \times N_{\text{dict}}}$ is compressed into a low-dimensional temporal subspace using singular value decomposition (SVD) \cite{McGivney2014}:
\[
D = U S V' \quad \rightarrow \quad D_{\text{SVD}} = D V_{\text{PC}}
\]
The compression matrix $V_{\text{PC}} \in \mathbb{C}^{N_R \times N_{\text{PC}}}$ contains the leading $N_{\text{PC}}$ principal components that preserve 99.99\% of the dictionary energy. The same matrix is applied to compress the raw k-space data along the TR dimension, and the resulting subspace data are reconstructed using a standard NUFFT operator \cite{Fessler2003}. Dictionary matching is then performed voxel-by-voxel in the temporal subspace.

In the second approach, the subspace coefficient maps are obtained using an iterative low-rank reconstruction \cite{Hamilton2019}, by solving a minimization problem:
\[
\min_x \| y - Sx \| + \sum_i \lambda_i |R_i x|
\]
Here, the signal operator $S = PFC$ includes the coil operator $C$, the NUFFT operator $F$ and the acquisition mask $P$. The sum $\lambda_i |R_i x|$ consists of multiple L1-normalized penalty terms (spatial total variation, locally low rank and wavelet regularization) with their corresponding weights $\lambda_i$. To support efficient reconstruction, the pipeline includes optional coil compression, as well as ROVIR-based suppression of signals outside the desired field of view \cite{Kim2021}. Coil sensitivity maps are estimated from the compressed multi-channel data using ESPIRiT \cite{Uecker2014}. The minimization problem is solved using nonlinear Conjugate Gradient Descent.

A brute-force NUFFT reconstruction followed by voxelwise dictionary matching, as described in Ma et al. \cite{Ma2013}, remains available in the toolbox for completeness but was not used in this study.

\subsubsection*{Storing and exchanging Pulseq metadata}
A general limitation of Pulseq - not specific to MRF - is the absence of a standardized mechanism for storing metadata. Essential sequence parameters such as preparation timings and k-space trajectories are implicitly defined in the .seq file but not explicitly embedded in the raw data exported by the scanner. This hampers reproducibility, automated reconstruction, and cross-site comparison. OpenMRF addresses this by introducing a metadata-handling concept. Metadata generated during sequence creation are stored in a structured, user-specific and time-stamped format, and the Pulseq MD5 hash provides a unique identifier connecting raw data to the corresponding backup. This enables automatic retrieval of all relevant sequence parameters during reconstruction and supports reproducible Pulseq-based studies across sites and vendors. Metadata retrieval is fully automated and does not require user input, eliminating a potential error source.

\subsection*{II. Implementation of example sequences}

To evaluate the flexibility and functionality of OpenMRF, three representative MRF protocols were implemented for brain, abdominal and cardiac applications. All sequences are openly available in the GitHub repository and the Pulseq sequence diagrams can be found in the Supplementary Material (\hyperref[sfig:irfisp-seq]{Suppl. Fig. 1-5}).

The first protocol is based on the IR-FISP MRF design of Jiang et al. \cite{Jiang2015} and features an adiabatic inversion pulse followed by 1000 variable flip-angle excitations with a FISP-type spiral readout. Flip angles follow sinusoidal modulation with a maximum of 70°, and repetition times are varied pseudorandomly. A variable-density spiral-out trajectory with 24/48-fold undersampling in k-space center/periphery was used, and gradient spoiling was applied along the slice-select direction.

Two additional cardiac MRF (cMRF) sequences were implemented featuring ECG triggering and segmented acquisitions. Each sequence consists of 16 acquisition windows with 48 spiral readouts per segment, with variable excitation flip angles between 5° and 15° and minimized TR. The newly released soft delay feature of Pulseq was used to adjust the acquisition window to diastole depending on the individual subject heart rate. The first cMRF sequence targets $T_1$ and $T_2$ mapping and follows the design of Hamilton et al. \cite{Hamilton2018}, including adiabatic inversions and $T_2$-preparation modules with BIR-4 rotations and composite refocusing pulses. The second variant additionally includes balanced spin-lock preparations for $T_{1\rho}$ encoding.

For 1.5\,T and 3\,T experiments, near-identical .seq files were used except for adjustments to the fat-saturation frequency. On the 0.55\,T system, the same sequence structures were maintained with three modifications to satisfy hardware limits: reduced spatial resolution, adiabatic pulses with lower peak amplitude, and a reduced spin-lock frequency (100\,Hz instead of 300\,Hz). For numerical simulations, phantom studies and abdominal acquisitions, ECG triggering was disabled, instead the acquisition windows were manually placed to emulate a heart rate of 60\,bpm. All sequence parameters, including flip-angle trains, preparation modules, and trajectory definitions, are documented in the GitHub repository.

\subsection*{III. Bloch simulations and digital phantom tests}

To validate the accuracy of the OpenMRF accelerated dictionary creation pipeline, a series of numerical experiments was performed using the example sequences described above. First, each .seq file was parsed and converted into simulation instructions suitable for the brute-force Bloch simulator. Pre-simulations were then conducted to characterize slice-profile effects and relaxation during adiabatic pulses, which are required for the accelerated simulation mode. The accelerated simulations were subsequently compared with brute-force references.

To assess the full end-to-end MRF workflow, a numerical phantom experiment was performed. A 2D digital phantom was populated with $T_1$ and $T_2$ values matching the ISMRM/NIST system phantom MnCl$_2$ insert specifications (inserts 1–14 at 1.5\,T). Synthetic MRI data were generated using the full acquisition model, including the spiral k-space trajectory. Realistic coil sensitivity maps, derived from previously conducted scans with a 20-channel head/neck array, were used to emulate multi-coil encoding. The resulting multi-coil non-Cartesian data were processed using the OpenMRF reconstruction pipeline, including subspace compression, low-rank reconstruction, and dictionary-based parameter matching. Estimated $T_1$ and $T_2$ maps were compared with ground-truth values to assess overall accuracy and to verify correct interaction between all modules of the framework.

\subsection*{IV. Phantom study}

To validate OpenMRF across multiple field strengths and platforms, the example sequences were executed on clinical Siemens systems located at three sites: University of Michigan Hospitals, Ann Arbor (Free.Max 0.55\,T, Sola 1.5\,T, Vida 3\,T), University Hospital Würzburg (Avanto 1.5\,T, Prisma 3\,T), and University Medical Center Freiburg (Aera 1.5\,T, Cima.X 3\,T). In addition, .seq files were compiled for a GE Signa Premier 3\,T system at the University of Michigan \cite{Nielsen2018} and for a United Imaging Healthcare 3\,T uMR790 system located in Houston, Texas. Sequence design on the GE and United Imaging platforms preserved all global timings and preparation modules while adapting gradient and RF raster constraints.

All measurements were performed using a standardized protocol. The ISMRM/NIST MRI system phantom \cite{Stupic2021} was positioned in a dedicated head coil with the MnCl$_2$ layer aligned coronally. After slice planning, the MnCl$_2$ slice was shimmed using the vendor’s standard routines. Reference measurements for $T_1$ and $T_2$ estimation were obtained using single-slice inversion-recovery spin-echo and spin-echo protocols, employing variable inversion delays and echo times, respectively. For $T_{1\rho}$, an additional Pulseq-based single-parameter mapping protocol (included in OpenMRF) was performed using the same spin-lock preparation module as implemented in the cMRF sequence.

To characterize B0 and $B_1^+$ inhomogeneities across the different systems, a WASABI sequence \cite{Schuenke2017b} was executed on each scanner following the MRF protocol. Subsequently, a spherical calibration phantom was used to measure the actual spiral k-space trajectories corresponding to the example sequences, which were used for the reconstruction.

\subsection*{V. \textit{In vivo} validation}

To evaluate the applicability of OpenMRF across different anatomical regions and field strengths, representative \textit{in vivo} acquisitions were performed in healthy volunteers on Siemens platforms at the University of Michigan. Three protocols were tested:
\begin{itemize}
\item Abdominal MRF at 0.55\,T using the cardiac MRF sequence without ECG triggering, targeting the liver and including $T_1$ and $T_2$ preparation modules.
\item Cardiac MRF at 1.5\,T using the full cMRF sequence with ECG triggering, soft-delay adjustment, and combined $T_1$, $T_2$, and $T_{1\rho}$ encoding.
\item Brain MRF at 3.0\,T using the IR-FISP protocol.
\end{itemize}

The study was approved by the University of Michigan Institutional Review Board, and written informed consent was obtained from all participants.

\begin{figure}[p]
    \centering
    \includegraphics[width=\textwidth]{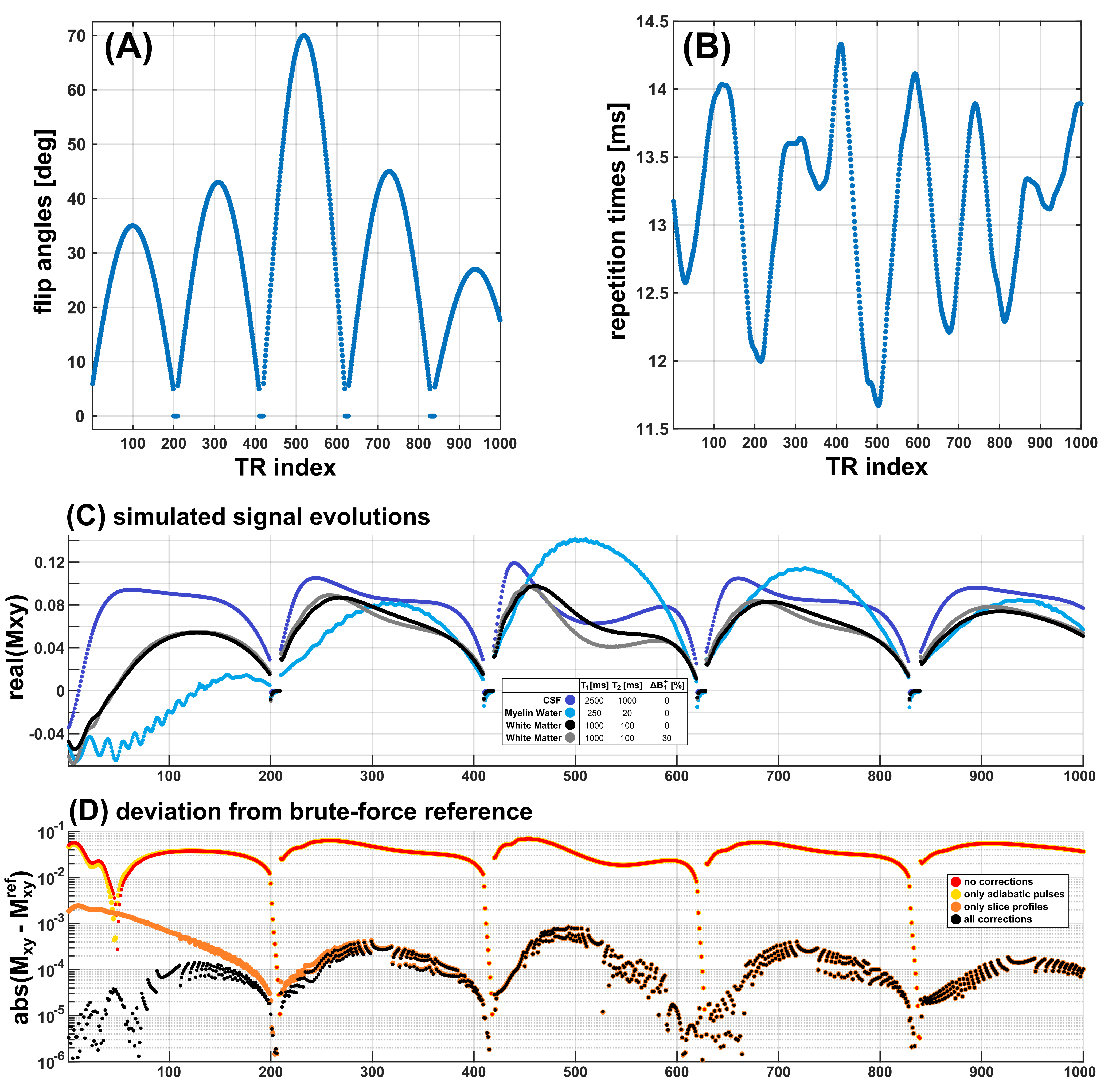}
    \caption{Sequence parameters and Bloch-simulated signal evolutions for the IR-FISP MRF experiment. (A) Variable flip-angle pattern and (B) variable repetition times used in the IR-FISP acquisition \cite{Jiang2015}. (C) Bloch-simulated transverse magnetization (real component of Mxy) for representative tissue parameter combinations spanning CSF, myelin water, and white matter with different $T_1$, $T_2$, and $B_1^+$ values; $B_1^+$ deviations clearly alter the signal evolution. (D) Deviation of simulated fingerprints from a brute-force reference solution, illustrating the impact of neglecting slice-profile effects and relaxation during adiabatic pulses. After applying the proposed corrections, the average deviation from the computationally costly brute-force method is $0.0121 \pm 0.0148$ \% (relative to $M_0 = 1$).}
    \label{fig:irfisp-sim}
\end{figure}

\begin{figure}[p]
    \centering
    \includegraphics[width=\textwidth]{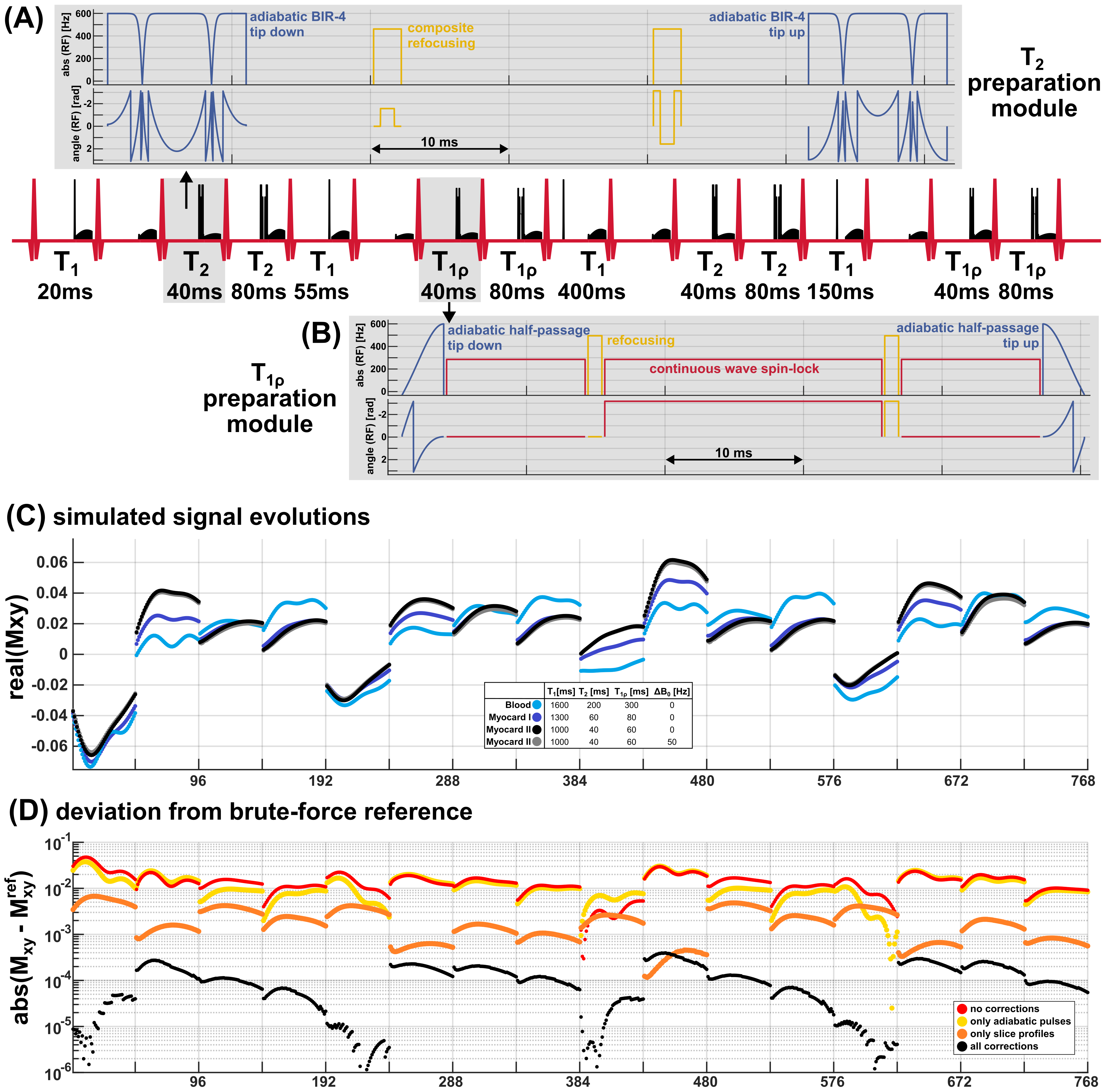}
    \caption{Preparation modules and Bloch-simulated signal evolutions for the $T_1$-$T_2$-$T_{1\rho}$ cMRF sequence. (A) $T_2$ preparation module (TE = 40\,ms) including adiabatic BIR-4 and composite refocusing pulses (B) $T_{1\rho}$ preparation module (tSL = 40\,ms) based on the balanced spin-lock approach \cite{Gram2021} (C) Bloch-simulated transverse magnetization (real component of Mxy) for representative tissue parameter combinations spanning different myocardial tissues and blood with different $T_1$, $T_2$, $T_{1\rho}$ and $\Delta$B0 values; $\Delta$B0 deviations slightly alter the signal response. (D) Deviation of simulated fingerprints from a brute-force reference simulation, illustrating the impact of neglecting slice-profile effects and relaxation during adiabatic pulses. After applying the proposed corrections, the average deviation is $0.0112 \pm 0.0100$ \%.}
    \label{fig:cmrf-sim}
\end{figure}

\newpage
\section{Results}

\subsection*{Bloch simulations and digital phantom validation}

\hyperref[fig:irfisp-sim]{Figure 2} shows the sequence parameters of the IR-FISP MRF sequence and simulated fingerprints for different brain tissues: CSF, white matter, and myelin water. Additionally, the simulated white matter fingerprint assuming a +30\% $B_1^+$ deviation is shown, demonstrating substantial deviations from the case of perfect B1 homogeneity. \hyperref[fig:irfisp-sim]{Figure 2D} illustrates the deviation of a simulated fingerprint from the brute-force reference. Only when all corrections (including slice-profile effects and relaxation during adiabatic pulses) are included, the deviations become negligible, with the slice-profile effect being the dominant contributor to the otherwise observed deviations. \hyperref[fig:cmrf-sim]{Figure 3} shows the same comparison for a $T_1$-$T_2$-$T_{1\rho}$ cMRF sequence. Here, relaxation effects during adiabatic pulses are even more important due to the large number of adiabatic modules with long pulse durations. Both the IR-FISP and cMRF sequences are modelled efficiently, with low residual deviations, using the accelerated Bloch-isochromat approach. The average deviations are $0.0121 \pm 0.0148$ \% (IR-FISP) and $0.0112 \pm 0.0100$ \% (cMRF). On a desktop PC with an AMD Ryzen 9 7950X CPU, the simulation time was 2.39\,s (brute force, cMRF) and 36.6\,ms (accelerated, cMRF) per dictionary entry.

\begin{figure}[p]
    \centering
    \includegraphics[width=\textwidth]{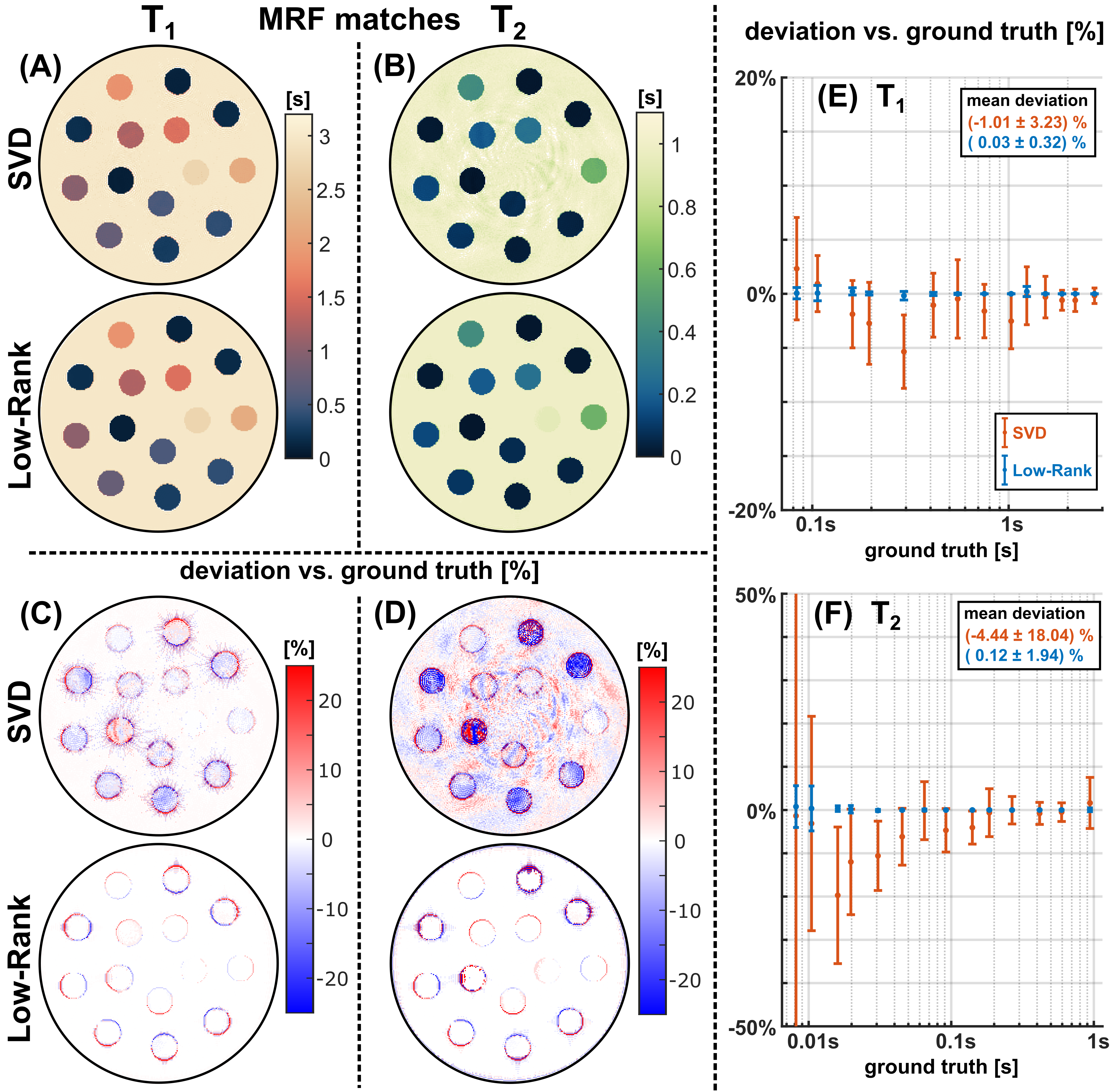}
    \caption{Digital phantom-based comparison of SVD and low-rank reconstruction for the IR-FISP MRF sequence. A-B) Quantitative $T_1$ and $T_2$ maps estimated using the SVD and low-rank approaches for the digital phantom, showing visually improved homogeneity and reduced noise with the low-rank method. C-D) Pixelwise deviation maps relative to ground-truth values, demonstrating substantially smaller errors for the low-rank reconstruction. E-F) Quantitative deviation plots across phantom inserts for $T_1$ and $T_2$, with mean deviations of $-1.01 \pm 3.33$ \% (SVD) and $0.03 \pm 0.32$ \% (low-rank) for $T_1$, and $-4.44 \pm 18.04$ \% (SVD) and $0.12 \pm 1.94$ \% (low-rank) for $T_2$. The corresponding simulation results for the cMRF variants are included in the Supplementary Material.}
    \label{fig:digital-phantom}
\end{figure}

The results of the validation in the numerical phantom are shown in \hyperref[fig:digital-phantom]{Figure 4}. Although parameter maps based on SVD-compressed subspace reconstruction demonstrate good image quality, residual aliasing artifacts are visually apparent in both the $T_1$ and $T_2$ maps (\hyperref[fig:digital-phantom]{Fig. 4A-B}). In the low-rank reconstructed maps, they are noticeably reduced. Deviations from the ground truth are small throughout the entire phantom (\hyperref[fig:digital-phantom]{Fig. 4C-F}). Using basic SVD reconstruction, the deviations are $-1.01 \pm 3.33$\% ($T_1$) and $-4.44 \pm 18.04$\% ($T_2$). Low-rank reconstruction provides improved accuracy and precision at $0.03 \pm 0.32$\% ($T_1$) and $0.12 \pm 1.94$\% ($T_2$). An analogous evaluation of the two cMRF variants can be found in the Supplementary Material (\hyperref[sfig:digital-tt-cmrf]{Suppl. Fig. 6-7}).

\subsection*{Phantom validation}

\begin{figure}[p]
    \centering
    \includegraphics[width=\textwidth]{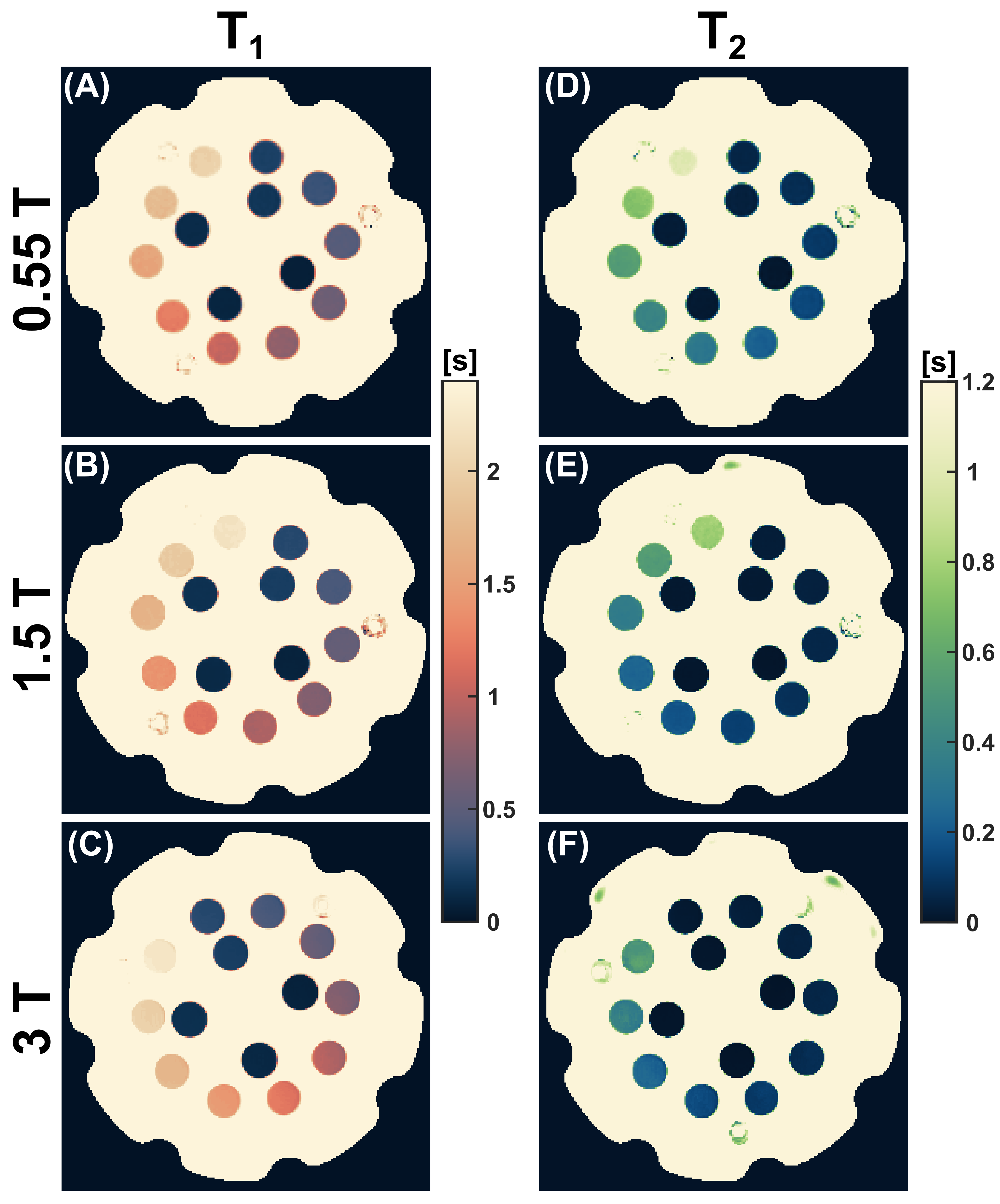}
    \caption{Multi-field-strength validation of the IR-FISP MRF sequence in the ISMRM/NIST system phantom acquired at the University of Michigan using Siemens Free.Max (0.55\,T), Sola (1.5\,T), and Vida (3\,T) scanners. All datasets were reconstructed using the low-rank subspace method. The dictionary was generated with $T_1$ values ranging from 10\,ms to 4\,s and $T_2$ values from 1\,ms to 3\,s in 2.5\% increments. Combinations with $T_1$ $<$ $T_2$ were removed. For the 3\,T measurements, $B_1^+$ deviations were additionally modeled. A--C) Reconstructed $T_1$ maps at 0.55\,T, 1.5\,T, and 3\,T. D--F) Corresponding $T_2$ maps.}
    \label{fig:phantom-irfisp}
\end{figure}

\begin{figure}[p]
    \centering
    \includegraphics[width=\textwidth]{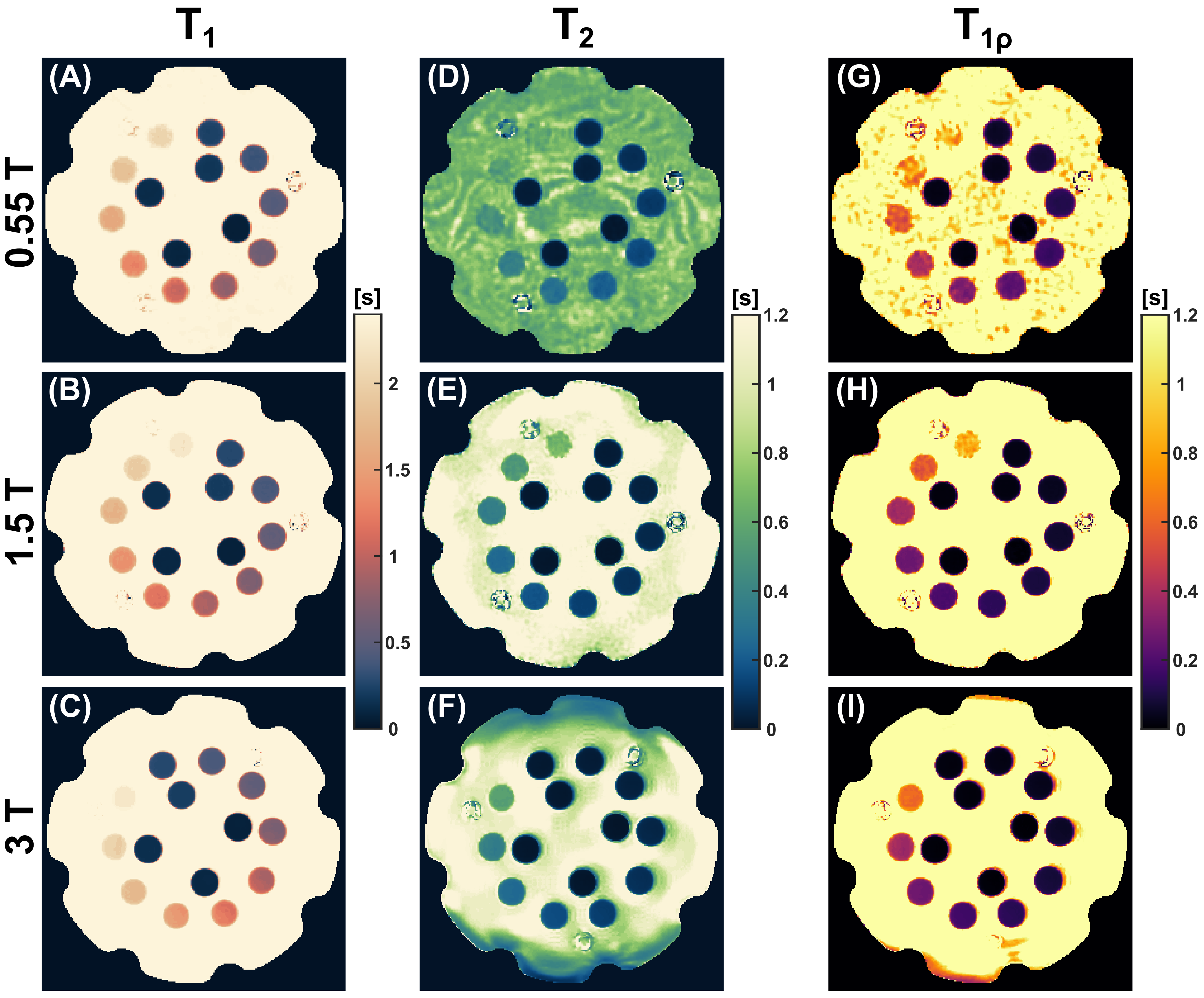}
    \caption{Multi-field-strength validation of the $T_1$-$T_2$-$T_{1\rho}$ cardiac MRF (cMRF) sequence in the ISMRM/NIST system phantom, acquired using Siemens Free.Max (0.55\,T), Sola (1.5\,T), and Vida (3\,T) scanners. All datasets were reconstructed using the low-rank subspace method. The dictionary was generated with $T_1$ values ranging from 10\,ms to 4\,s in 5\% steps, and $T_2$ and $T_{1\rho}$ values ranging from 1\,ms to 3\,s in 5\% steps. Only physically relevant combinations were used ($T_2$ $<$ $T_1$, $T_{1\rho}$ $<$ $T_1$, and $T_2$ $<$ $T_{1\rho}$). A--C) Reconstructed $T_1$ maps, D--F) $T_2$ maps, and G--I) $T_{1\rho}$ maps at 0.55\,T, 1.5\,T, and 3\,T, respectively. Simultaneous estimation of three relaxation parameters ($T_1$, $T_2$, $T_{1\rho}$) was successfully achieved. $T_{1\rho}$ values are slightly higher than $T_2$ due to small dispersion effects. Image quality is generally high across field strengths, though increased noise is visible in the $T_{1\rho}$ map at 0.55\,T and slight banding artifacts appear in the $T_2$ map at 3\,T.}
    \label{fig:phantom-cmrf}
\end{figure}

\begin{figure}[p]
    \centering
    \includegraphics[width=\textwidth]{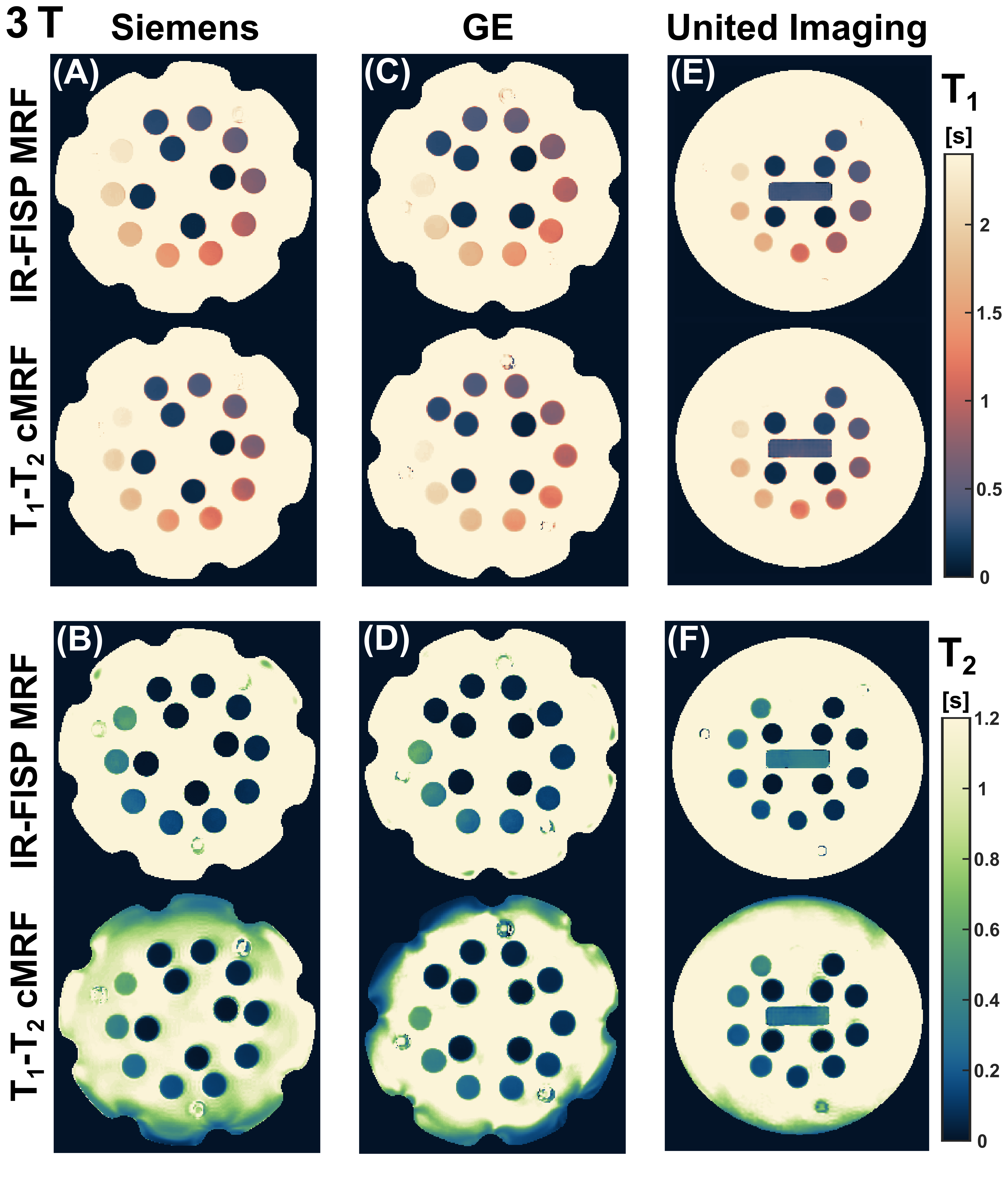}
    \caption{Multi-vendor comparison of quantitative parameter maps obtained with IR-FISP MRF and $T_1$-$T_2$ cMRF at 3\,T using the OpenMRF framework. Measurements were performed on three platforms: Siemens Vida (A,B), GE Signa (C,D), and United Imaging Healthcare uMR790 (E,F). Siemens and GE scans used the same NIST phantom. All sequences were implemented using nearly identical Pulseq .seq files, with only minor adjustments regarding raster timings required to accommodate vendor-specific hardware constraints. Data were reconstructed and processed solely based on the OpenMRF pipelines, including dictionary simulation, low-rank reconstruction and pattern matching. The results demonstrate good overall agreement in $T_1$ and $T_2$ estimates across vendors, with consistent contrast between phantom inserts. Residual spatial variations, particularly in $T_2$ maps, reflect differences in system characteristics.}
    \label{fig:vendor-comparison}
\end{figure}

The results of the multi-site phantom study are shown in \hyperref[fig:phantom-irfisp]{Figures 5-9}. Representative parameter maps acquired in Michigan using the IR-FISP sequence and reconstructed using the low-rank approach are provided in \hyperref[fig:phantom-irfisp]{Figure 5}, demonstrating high image quality across all three field strengths (0.55\,T, 1.5\,T, 3\,T). The corresponding measurements using the $T_1$-$T_2$-$T_{1\rho}$ cMRF variant (\hyperref[fig:phantom-cmrf]{Fig. 6}) also exhibit high quality, with only minor banding artifacts in the 3\,T $T_2$ map and increased noise in the 0.55\,T $T_{1\rho}$ map. For 3\,T modelling, the $B_1^+$ effect was included as an additional matched parameter due to the $\pm$30\% transmit-field variation commonly observed at this field strength. A direct comparison between a WASABI-measured $B_1^+$ map and a $B_1^+$ map estimated via $T_1$-$T_2$ cMRF matching shows high structural similarity (\hyperref[sfig:b1-wasabi]{Suppl. Fig. 8-9}). \hyperref[fig:vendor-comparison]{Figure 7} illustrates the direct comparison between 3\,T tests performed at Siemens, GE and United Imaging systems. For all vendors, high image quality was obtained with similar residual artifacts for $T_2$ maps.

Measured spiral-out trajectories used for reconstruction are depicted in \hyperref[sfig:traj-comparison]{Supplementary Figures 10-11}. The method of Robison et al. was successfully applied at all field strengths and sites, with only small deviations from the Pulseq-predicted trajectories. The average deviation relative to $k_{\max}$ was 0.16 \%.

\begin{figure}[p]
    \centering
    \includegraphics[width=\textwidth]{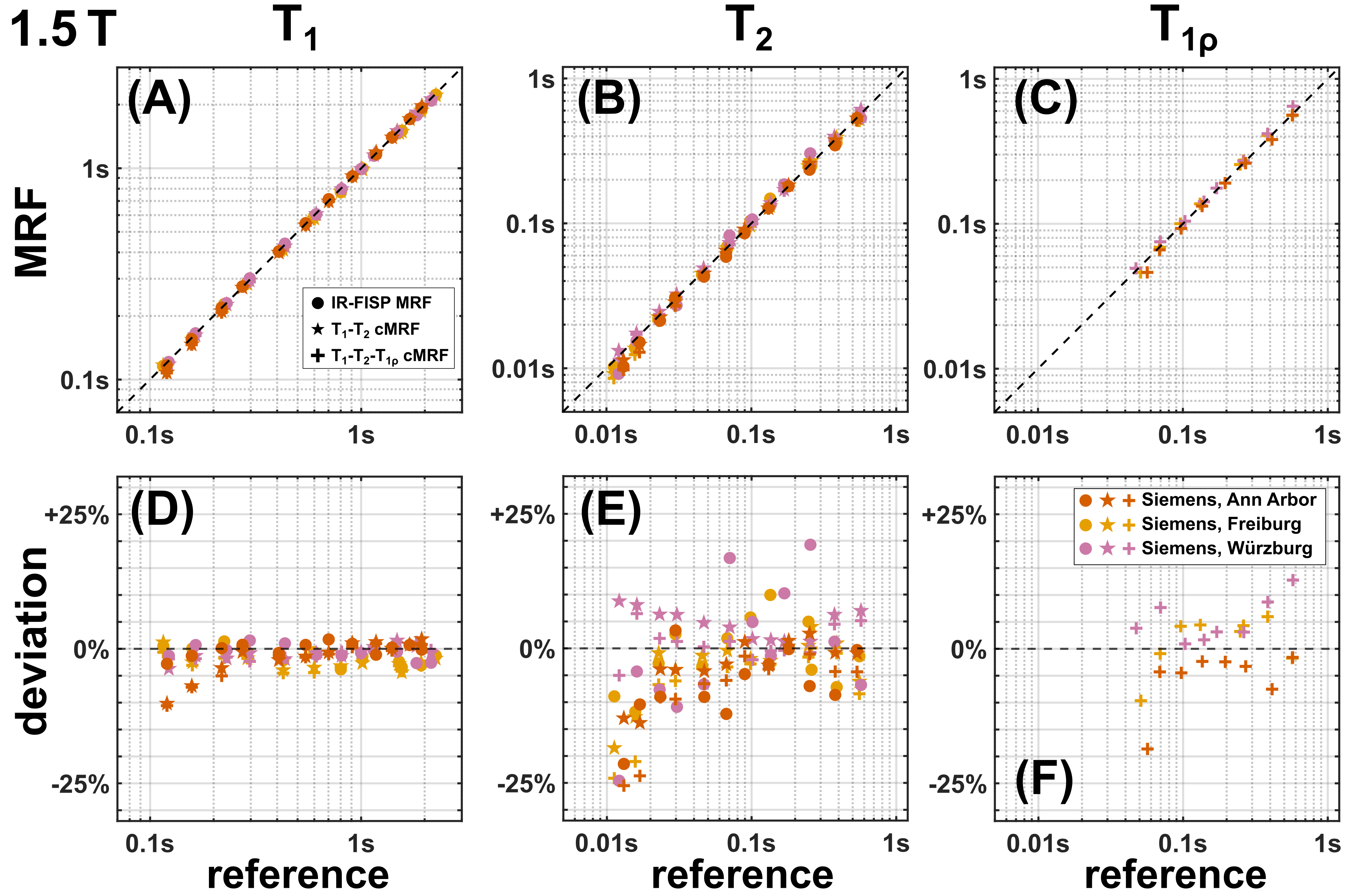}
    \caption{Multi-site comparison of $T_1$, $T_2$, and $T_{1\rho}$ estimates obtained with three MRF protocols (IR-FISP MRF, $T_1$-$T_2$ cMRF, and $T_1$-$T_2$-$T_{1\rho}$ cMRF) at 1.5\,T in the ISMRM/NIST system phantom. Data were acquired at three sites (W\"urzburg, Freiburg, and Michigan) using Siemens Avanto, Aera, and Sola scanners. A--C) Estimated relaxation times plotted against the gold-standard reference values for $T_1$, $T_2$, and $T_{1\rho}$. D--F) Percentage deviations relative to the reference measurements, indicating consistently small deviations for $T_1$ and $T_{1\rho}$ and increased variability in $T_2$ across scanners and protocols. These results confirm robust cross-site reproducibility of the proposed OpenMRF protocols at 1.5\,T.}
    \label{fig:phantom-15t}
\end{figure}

\begin{figure}[p]
    \centering
    \includegraphics[width=\textwidth]{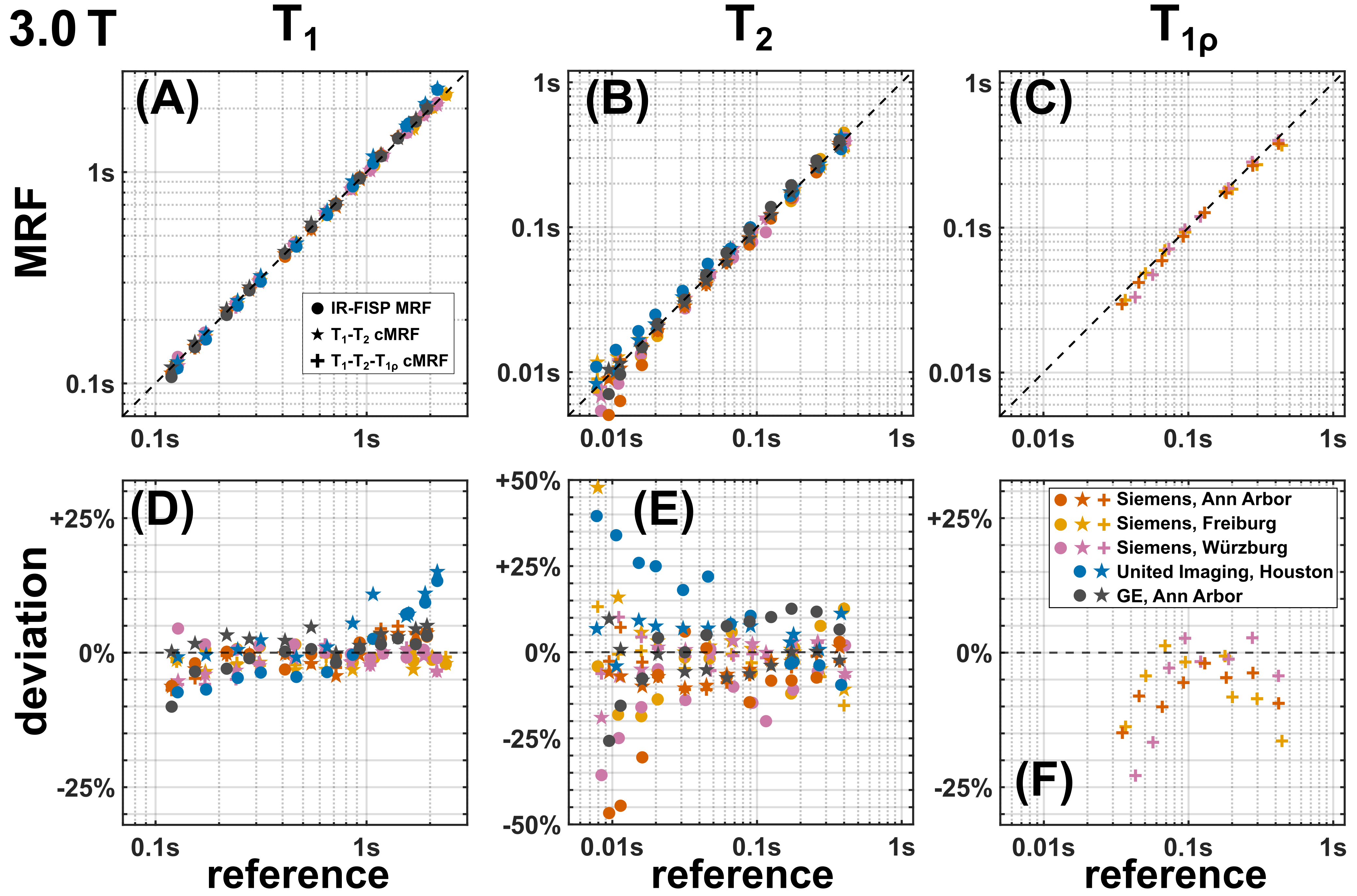}
    \caption{Multi-site comparison of $T_1$, $T_2$, and $T_{1\rho}$ estimates obtained with three MRF protocols (IR-FISP MRF, $T_1$-$T_2$ cMRF, and $T_1$-$T_2$-$T_{1\rho}$ cMRF) at 3\,T in the ISMRM/NIST system phantom. Data were acquired at four sites (W\"urzburg, Freiburg, Michigan, Houston) using Siemens Prisma, Cima.X and Vida, as well as GE Signa and United Imaging uMR790 systems. A--C) Estimated relaxation times plotted against the gold-standard reference values for $T_1$, $T_2$, and $T_{1\rho}$. D--F) Percentage deviations relative to the reference values, showing small deviations for $T_1$ and $T_{1\rho}$, and increased variability in $T_2$ across scanners and protocols. These results confirm robust cross-site reproducibility of the proposed OpenMRF protocols at 3\,T, with only one outlier observed in $T_1$ estimation using the $T_1$-$T_2$- $T_{1\rho}$ sequence.}
    \label{fig:phantom-3t}
\end{figure}

\hyperref[fig:phantom-15t]{Figures 8-9} present a comparison of the $T_1$, $T_2$, and $T_{1\rho}$ values of the MnCl$_2$ inserts determined with MRF against the gold standard reference values. At 0.55\,T (\hyperref[sfig:phantom-055t]{Suppl. Fig. 12}), the mean deviation across all sequences was $2.77 \pm 2.38$ \% for $T_1$ and $-0.41 \pm 8.35$ \% for $T_2$, with the cMRF variants showing slightly larger deviations. For the averaged deviations, only the MnCl$_2$ inserts \#3-12 were evaluated, representing a clinically relevant range of tissue relaxation times. $T_{1\rho}$ was evaluated only for inserts \#6-12, as the SL reference sequence did not provide reliable results for low relaxation values. In this subset, the deviations were $-9.7 \pm 4.5$ \%. Note that 0.55\,T data were acquired at a single site (Michigan), whereas 1.5\,T (\hyperref[fig:phantom-15t]{Fig. 8}) and 3\,T (\hyperref[fig:phantom-3t]{Fig. 9}) were validated at three and four sites each. The averaged deviations across all vendors, sites and sequences for 1.5\,T were $-1.29 \pm 1.86$\% ($T_1$), $-1.66 \pm 6.67$\% ($T_2$), and $-0.11 \pm 6.44$ \% ($T_{1\rho}$). At 3\,T, the mean deviations were $0.11 \pm 3.12$ \% ($T_1$), $-1.63 \pm 10.01$ \% ($T_2$), and $-5.93 \pm 6.70$\% ($T_{1\rho}$). \hyperref[sfig:repeatability]{Supplementary Figure 13} further reports repeatability for IR-FISP MRF. Across five repeated scans of the IR-FISP MRF, the averaged variability was 0.23 \% ($T_1$) and 0.51 \% ($T_2$) at 1.5\,T and 0.39 \% ($T_1$) and 1.6 \% ($T_2$) at 3.0\,T.

\subsection*{\textit{In vivo} application}

\begin{figure}[p]
    \centering
    \includegraphics[width=\textwidth]{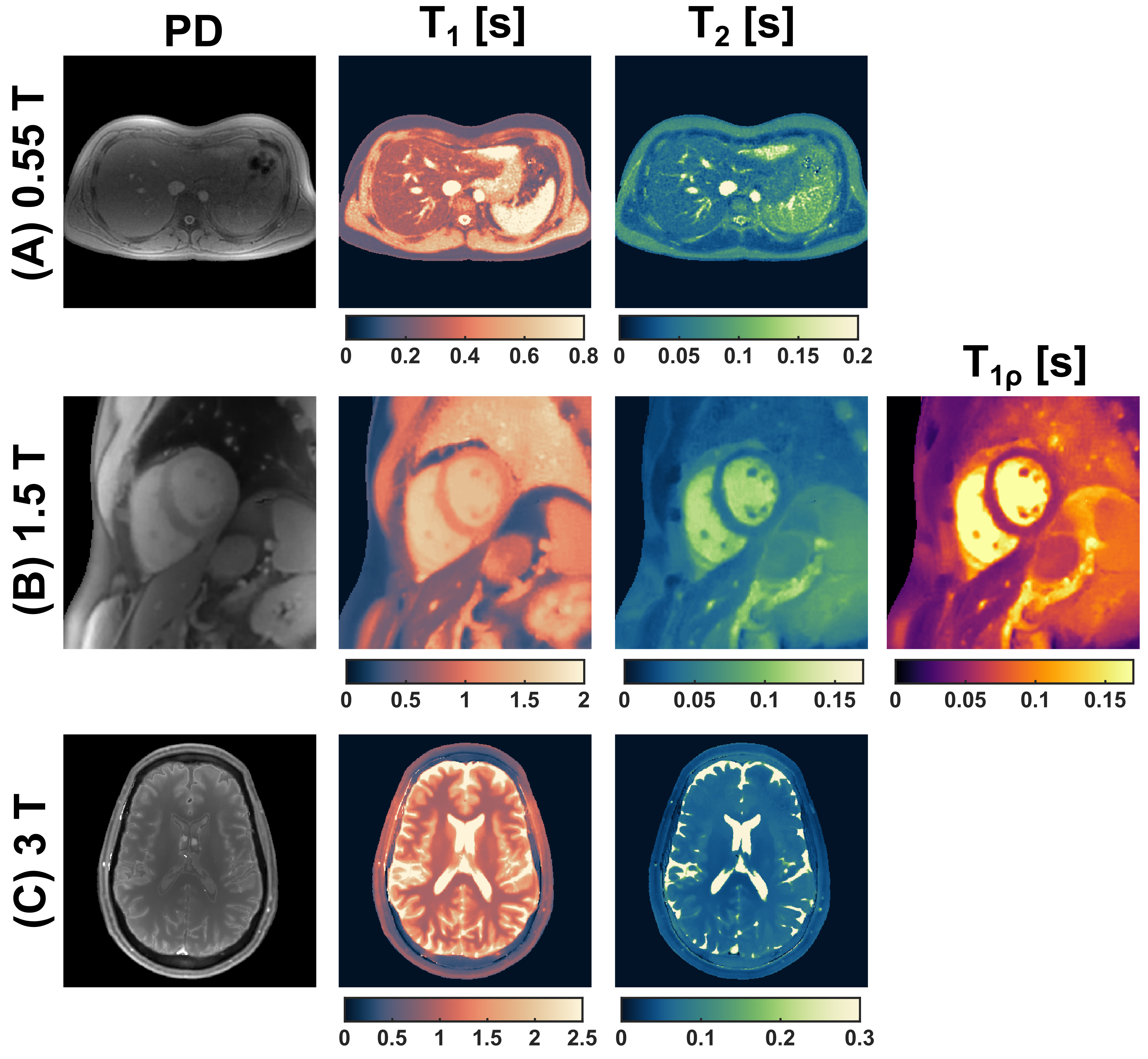}
    \caption{\textit{In vivo} demonstration of the OpenMRF framework across multiple field strengths. A) Liver imaging at 0.55\,T using the $T_1$-$T_2$ cMRF sequence with a fixed segment duration of 1\,s. B) Cardiac short-axis imaging at 1.5\,T using the $T_1$-$T_2$-$T_{1\rho}$ cMRF sequence with prospective R-wave triggering. The trigger timings were automatically parsed to the dictionary simulation, and the cardiac acquisition window was adapted using the soft-delay feature. C) Brain imaging at 3\,T using the IR-FISP MRF sequence. Across all field strengths and anatomical regions, the estimated $T_1$, $T_2$, $T_{1\rho}$, and proton-density maps exhibit high image quality and estimated parameters are generally within expected ranges, though $T_2$ at 3\,T might be underestimated.}
    \label{fig:invivo}
\end{figure}

\hyperref[fig:invivo]{Figure 10} shows the results of the \textit{in vivo} example acquisitions. High-quality parameter maps were obtained at all three field strengths and for all anatomical applications. At 0.55\,T (liver), average relaxation times of $T_1 = 310.2 \pm 25.5$\,ms and $T_2 = 46.0 \pm 9.3$\,ms were observed. In the myocardium at 1.5\,T, the estimated values were $T_1 = 1129 \pm 20$\,ms, $T_2 = 36.0 \pm 1.3$\,ms and $T_{1\rho} = 51.0 \pm 1.5$\,ms. At 3\,T (brain), the relaxation times obtained for white matter were $T_1 = 939.2 \pm 7.1$\,ms and $T_2 = 43.4 \pm 1.5$\,ms, and for gray matter we observed $T_1 = 1363.4 \pm 47.0$\,ms and $T_2 = 59.1 \pm 2.1$\,ms. Cardiac MRF required prospective ECG triggering and dictionary generation based on exact trigger delays.

\newpage
\section{Discussion}

This study presents the first open-source, end-to-end framework for Magnetic Resonance Fingerprinting that fulfills all core requirements of modern MRF, from modular sequence implementation and automated dictionary simulation to advanced reconstruction and quantitative parameter estimation. The developed sequences and processing methods were validated using numerical simulations and a multi-site ISMRM/NIST system phantom study. The primary goal of this work was not to introduce or optimize a specific new MRF method, but to provide a general, extensible, and transparent framework that can serve as a foundation for future studies within the research community and to demonstrate multi-vendor applicability of standardized MRF. To support reproducibility and collaborative development, the complete source code is openly accessible on GitHub (github.com/OpenMRF).

\subsection*{Sequence implementation}

With the primary objective of demonstrating the validity and accuracy of our proposed framework, we focused on established MRF sequence designs commonly used for brain \cite{Jiang2015} and cardiac \cite{Hamilton2018} imaging. These sequences employ a FISP-type readout in conjunction with spiral k-space trajectories. Importantly, the modular implementation permits the integration of various other readout methods. For example, OpenMRF acquisitions with rosette k-space readouts have already been validated for cardiac imaging in \cite{KaplanISMRM2026} and for abdominal imaging in \cite{GrieslerISMRM2026b}, where they were combined with fat-fraction quantification \cite{Hernando2010, Liu2021}. For rosette acquisitions specifically, direct measurement of the k-space trajectory is feasible and advisable, as the repeated k-space center revisits can exacerbate trajectory deviations. In this work, trajectories were measured using the method described by Robison et al. \cite{Robison2019}. This method is robust but requires relatively lengthy calibration procedures since each interleaf is acquired individually. Furthermore, the calibration is specific for a single slice orientation, meaning any adjustment to the slice prescription requires repeating the calibration sequence, a notable practical constraint for clinical studies. A promising alternative is global characterization of the gradient system transfer function, which we plan to integrate in future releases \cite{Scholten2023}.

A current limitation of OpenMRF is that not all sequence modules are supported across all vendor platforms. In particular, $T_1$-$T_2$-$T_{1\rho}$ cMRF experiments could only be executed on Siemens systems, as long-duration spin-lock pulses are not yet supported by the current Pulseq interpreter implementations for GE and United Imaging.

\subsection*{Dictionary calculation}

A second key feature of OpenMRF is the fully automatic dictionary simulation derived directly from the Pulseq hardware instructions. Unlike previously published simulators, OpenMRF accurately models slice-profile effects, relaxation during adiabatic pulses, and automatically detects spin-lock conditions, switching to a Bloch formulation that incorporates $T_{1\rho}$ and $T_{2\rho}$ relaxation as well as B0 and $B_1^+$ inhomogeneities. Diffusion can also be included via the apparent diffusion coefficient, although this feature has not yet been validated, and magnetization-transfer modeling is planned for future releases. An adiabatic $T_{1\rho}$ preparation has additionally been implemented and compared with the conventional continuous-wave approach \cite{GrieslerISMRM2026b, Coletti2023, Si2025}.

For the simulations presented in this work, we relied on the Bloch-isochromat approach because it accurately captures slice-profile effects. An EPG-based simulator is also available in the framework and accounts for relaxation during adiabatic pulses, yielding fingerprints that closely approximate Bloch-isochromat results when assuming an ideal slice profile in a reduced computation time. However, the EPG simulator, which is based on \cite{Weigel2015}, does not model slice-profile imperfections and $T_{2\rho}$ decay during spin-lock preparations. Dictionary generation is currently CPU-accelerated with parallelization across entries, and while the accelerated Bloch simulator performs well (even with 1000 isochromats), the computational burden becomes substantial for large dictionaries or high-dimensional parameter spaces. GPU-based acceleration is therefore a logical next step for future development.

\subsection*{Reconstruction}

For image reconstruction, we employed the iterative low-rank subspace method developed by Hamilton et al. \cite{Hamilton2019}. This approach is widely used and well established, and in our study, it provided robust performance both in the digital phantom validation and across experiments at multiple field strengths. Compared with the simpler NUFFT-based reconstruction in the SVD subspace, the low-rank method provided superior image quality and substantially improved accuracy. During implementation, particular effort was made to ensure that all post-processing steps operate solely on raw data generated within the OpenMRF framework. Noise prescans were integrated directly into the .seq files and executed before the first RF excitation, and the corresponding whitening matrix was computed using Cholesky factorization \cite{Kellman2005}. Coil sensitivity maps were estimated from the MRF dataset using ESPIRiT, eliminating the need for vendor-specific prescan procedures.

Despite its advantages, the iterative low-rank algorithm remains a computational bottleneck of the pipeline. On a desktop workstation, reconstruction of a 2D dataset required 19 minutes for 50 iterations, making hyper-parameter tuning challenging and 3D data processing time-consuming \cite{Stebani2025}. Analogous to the dictionary creation, future releases will therefore include a GPU-implemented reconstruction option.

\subsection*{Residual quantification errors in the multi-site phantom study}

The multi-site phantom experiments demonstrate that OpenMRF enables near-identical protocol execution across different scanners and field strengths, which is an essential prerequisite for future repeatability studies. Previous work has shown that MRF can achieve robust $T_1$ and $T_2$ estimation in standardized phantom settings \cite{Jiang2017, Lo2022, Statton2022}, and our results are consistent with these findings. Across all systems, $T_1$ deviations were generally smaller than $T_2$ deviations, in agreement with earlier reports \cite{Jiang2017}. Larger $T_2$ errors, particularly for IR-FISP at 3\,T, are expected because this sequence is more sensitive to B0 and $B_1^+$ inhomogeneities, whereas adiabatic preparation modules in the cMRF variants provide improved robustness. However, several limitations must be considered when interpreting the remaining discrepancies. First, no repeatability measurements over consecutive days were performed, preventing a clear separation of intra-scanner from inter-scanner variability. Second, each site used a different NIST phantom, introducing unavoidable variability related to manufacturing tolerances. Third, no temperature correction was applied, although temperature is known to influence $T_1$ and $T_2$ estimates in the NIST phantom \cite{Statton2022}. The latter two effects were mitigated by using individual ground truth characterizations instead of relying on the reported phantom relaxation times. Furthermore, the present study included a broader evaluation across seven Siemens platforms, whereas only a single system from GE and United Imaging was assessed. As a result, no robust conclusions regarding vendor-specific differences in quantitative measurements can be drawn at this stage. In addition, \textit{in vivo} validation was limited to one single site, and further multi-center \textit{in vivo} studies are required to more comprehensively assess robustness across platforms and clinical settings.

Despite these limitations, the ability to reproduce near-identical Pulseq-based MRF protocols across multiple field strengths and vendors represents a key step toward standardized multi-site MRF. Future work can now build upon this foundation to more rigorously investigate repeatability and reproducibility.

\subsection*{Additional framework capabilities}

In addition to the core MRF pipeline, OpenMRF provides several features that enhance reproducibility and support quantitative MRI experiments beyond fingerprinting. A key component is the Pulseq-specific metadata concept. This mechanism improves transparency in collaborative environments and enables reliable retrieval of sequence parameters during reconstruction. We apply this workflow routinely in our laboratories in Würzburg and Michigan and have found that it substantially reduces human error, improves reproducibility, and supports transparent documentation of experiments. We believe that standardized metadata handling should become an integral component of future Pulseq-based development, particularly for quantitative applications.

Beyond MRF, the framework includes a range of additional MRI tools. These encompass single-parameter spiral mapping sequences for $T_1$, $T_2$, and $T_{1\rho}$ that have already been applied in a clinical study for the assessment of age-related intervertebral disc degeneration \cite{Notni2025}, as well as recently developed methods for $T_{2\rho}$ and adiabatic $T_{1\rho}$ estimation \cite{GrieslerISMRM2026b, Gram2024}. The spin-lock package was developed based on our preclinical work \cite{Gram2021, Gram2022a, Gram2022b} and can also be used for the detection of biomagnetic fields \cite{Gram2022c, Albertova2024}.

\subsection*{Future development and extensions}

Several developments are currently underway to further expand the capabilities of OpenMRF. With the release of Pulseq version 1.5.1, rotation objects based on quaternions have been introduced, enabling more efficient .seq file compression since interleaves of spiral or rosette trajectories no longer need to be stored individually. Although the benefits for 2D sequences are modest, this concept greatly simplifies implementation and substantially reduces file size for 3D trajectories such as cones. In the latest release of OpenMRF, we therefore introduced rotation objects and integrated the time-optimized gradient waveform design based on the methods proposed by Lustig et al. \cite{Lustig2008, Vaziri2012}.

One important direction is the translation of key components into Python \cite{WangISMRM2026}. While Pulseq originated in MATLAB and sequence development remains most advanced there, dictionary simulation and MRF reconstruction are better suited for GPU acceleration when implemented in Python. For this reason, we plan to provide accelerated PyTorch-based versions of these modules. In addition, we intend to incorporate Deep Image Prior (DIP) reconstruction directly into OpenMRF. DIP has demonstrated improved image quality compared with conventional low-rank methods and has shown promise for accelerating cardiac imaging, particularly at lower field \cite{Hamilton2022, Hamilton2023}.

Finally, we aim to broaden the cross-vendor compatibility of the framework. In addition to the Siemens, United Imaging and GE validations presented in this study, we have successfully ported and simulated the demo sequences for Philips systems \cite{Roos2025}. Moreover, initial tests were conducted on a 5\,T United Imaging system. Here, IR-FISP and cMRF variants were tested in the ISMRM/NIST phantom, and an initial \textit{in vivo} proof-of-concept measurement has been demonstrated in the heart \cite{GramISMRM2026b}.

\newpage
\section{Conclusion}

In this work, we introduce OpenMRF, the first open-source, Pulseq-based framework that integrates all essential components of Magnetic Resonance Fingerprinting. Comprehensive validation through numerical simulations, multi-site phantom experiments, and representative \textit{in vivo} measurements demonstrated that the framework provides accurate simulations, robust reconstructions, and vendor-neutral quantitative mapping across field strengths. By combining modular sequence development, automatic dictionary generation, and advanced reconstruction methods within a unified workflow, OpenMRF establishes a transparent and reproducible foundation for quantitative MRI research. The framework is not tied to a specific MRF implementation and is intended as a general platform to support method development, comparison, and multi-site harmonization. Ongoing work focuses on GPU-accelerated simulation and reconstruction, advanced gradient waveform design, and broader multi-vendor compatibility. We believe that OpenMRF will facilitate collaborative research within the MRI community and contribute to more reproducible and transferable MRF studies in the future.

\section*{Author Contributions}

Study conception and study design: \textbf{T.G.}, \textbf{N.S.}, \textbf{M.G.}; data acquisition and ethics approval (Michigan): \textbf{T.G.}, \textbf{S.K.}, \textbf{N.S.}, \textbf{J.-F.N.}, \textbf{M.G.}; data acquisition (Würzburg): \textbf{T.W.}, \textbf{M.G.}; data acquisition (Freiburg): \textbf{Q.C.}, \textbf{X.W.}, \textbf{M.Z.}; data acquisition (Houston): \textbf{Z.Z.}, \textbf{Q.L.}, \textbf{P.M.}; data processing and analysis: \textbf{T.G.}, \textbf{M.G.}; development of the core OpenMRF framework: \textbf{T.G.}, \textbf{M.G.}; development of the low-rank reconstruction method: \textbf{J.I.H.}; development of foundational Pulseq packages or additional simulation and post-processing routines: \textbf{M.Z.}, \textbf{J.-F.N.}, \textbf{M.B.}, \textbf{P.A.}, \textbf{J.S.}, \textbf{I.A.}; guarantors of the integrity of the entire study: \textbf{N.S.}, \textbf{M.G.}; all authors reviewed the manuscript, contributed to its revision, and approved the final version.

\section*{Acknowledgments}
Work performed at the University of Michigan was supported by the National Institutes of Health (NIH/NHLBI HL163030, NIH R03 EB038095, NIH U24 NS120056, NIH R01 EB032378). Work performed at the University Medical Center Freiburg was supported by the National Institutes of Health (NIH R01 EB032378, NIH U24 NS120056), the European Union (MRITwins 101078393), EURAMET (22HLT02 A4IM), and the Deutsche Forschungsgemeinschaft (DFG, German Research Foundation; INST 39 1365-1). The Michigan Institute for Imaging Technology and Translation (MIITT) receives research support from Siemens Healthineers. M. Gram received funding from the International Society for Magnetic Resonance in Medicine (ISMRM) through the Research Exchange Grant Program.

\section*{Conflicts of Interest}
Jesse I Hamilton and Nicole Seiberlich receive research support from Siemens Healthineers and hold patents related to Magnetic Resonance Fingerprinting (MRF) technology. Qi Liu, Zhibo Zhu, and Peter Martin are employees of United Imaging Healthcare. All remaining authors declare no conflicts of interest relevant to this work.

\newpage
\bibliographystyle{unsrt}
\bibliography{mybib}

\setcounter{figure}{0}
\renewcommand{\figurename}{Supplementary Figure}

\newpage
\section*{Supplementary Material}
\FloatBarrier

\begin{figure} [h]
    \centering
    \includegraphics[width=\textwidth]{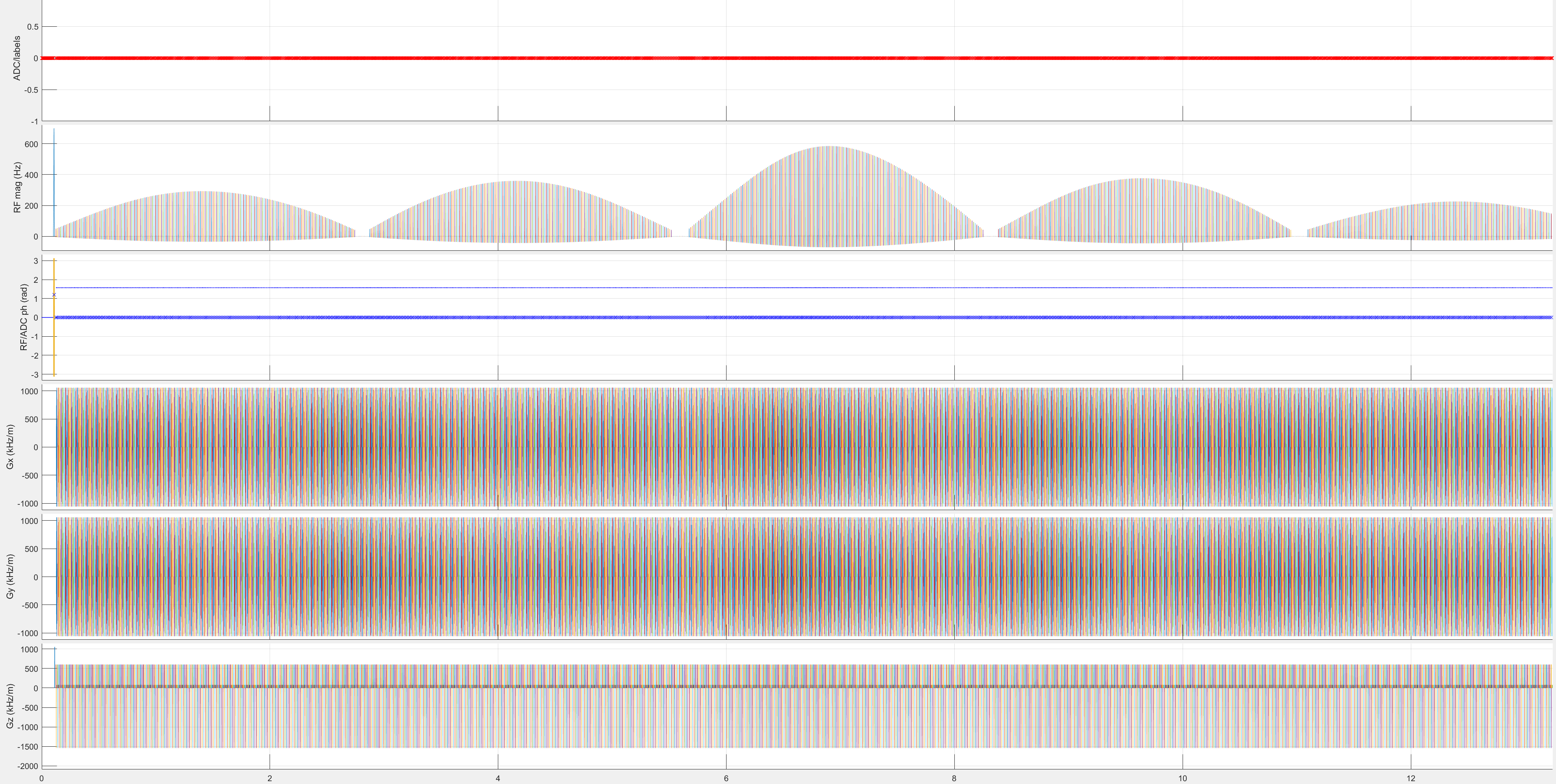}
    \caption{Pulseq sequence diagram of the IR-FISP MRF sequence. The implemented method is based on variable flip angles and repetition times following an adiabatic inversion.}
    \label{sfig:irfisp-seq}
\end{figure}
\clearpage

\begin{figure}[p]
    \centering
    \includegraphics[width=\textwidth]{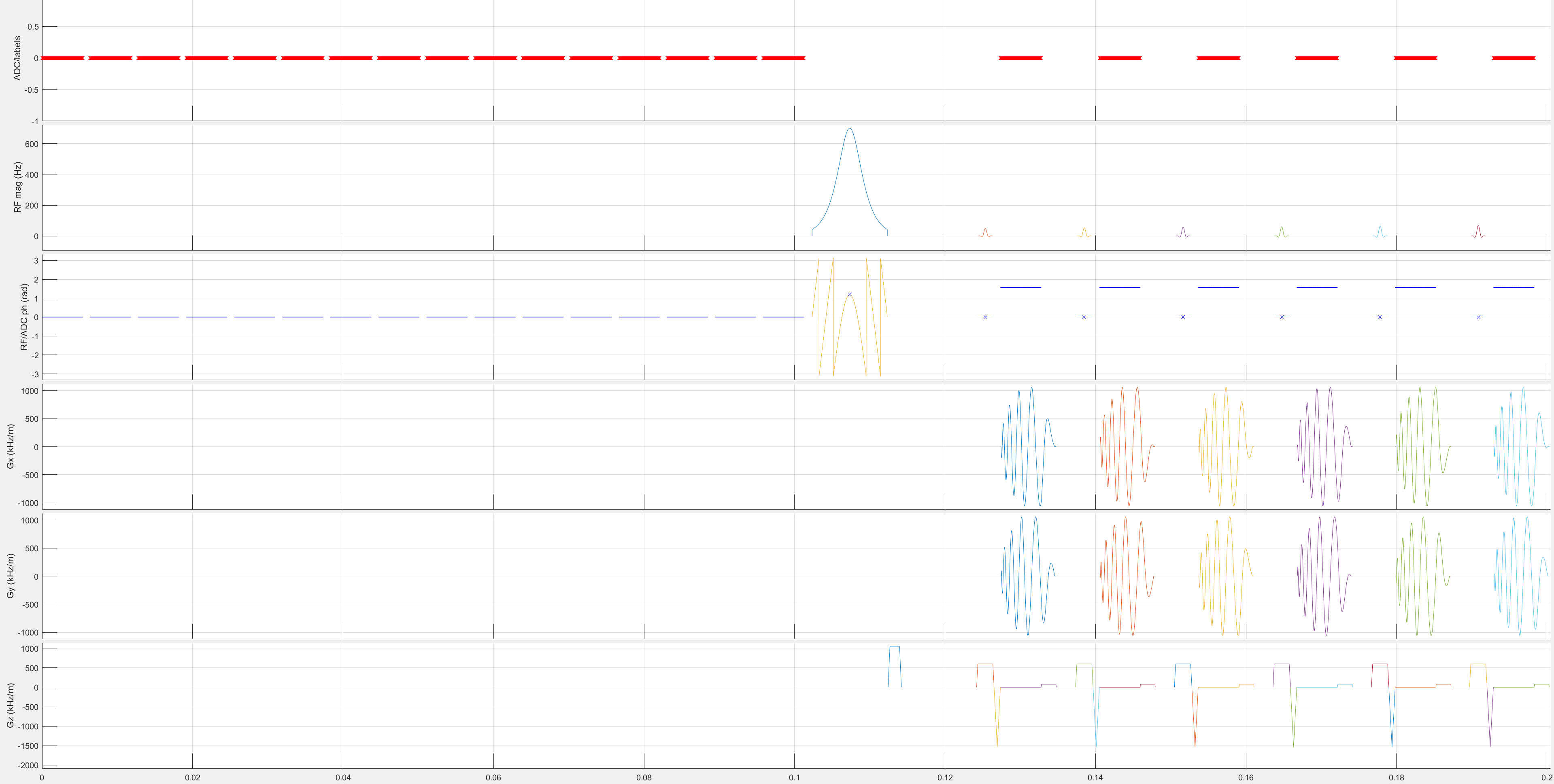}
    \caption{Enlarged section of the IR-FISP MRF sequence. The acquisition starts with multiple noise pre-scans, while the actual MRF sequence begins with an adiabatic inversion pulse (hyperbolic secant). The figure also depicts the first six spiral readouts.}
    \label{sfig:irfisp-detail}
\end{figure}
\clearpage

\begin{figure}[p]
    \centering
    \includegraphics[width=\textwidth]{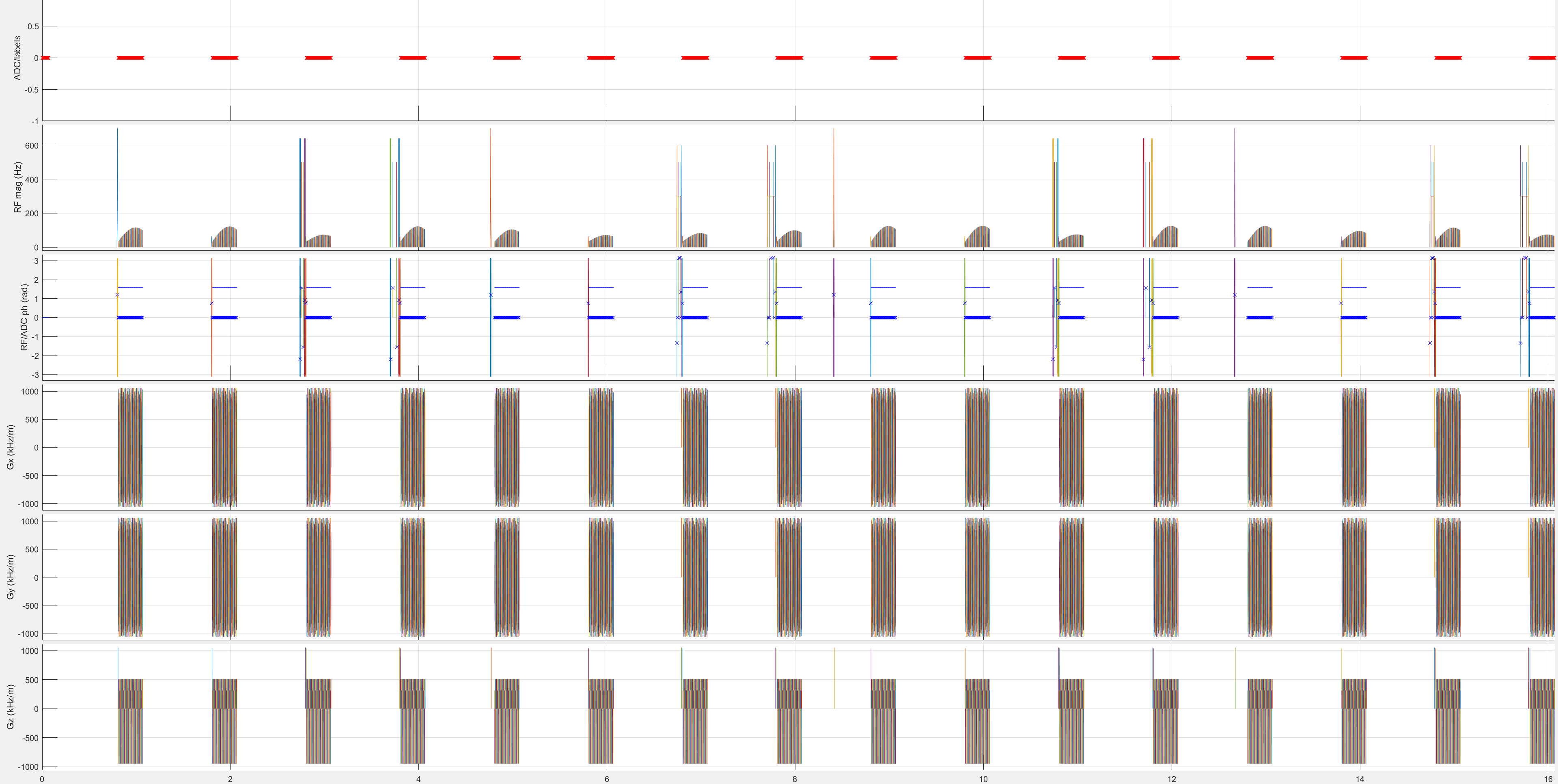}
    \caption{Pulseq sequence diagram of the $T_1$-$T_2$-$T_{1\rho}$ cMRF sequence. The method can be used for abdominal imaging, or for the heart by combining the protocol with prospective triggering. The sequence consists of 16 segments with different preparation modules, each followed by 48 spiral readouts.}
    \label{sfig:cmrf-seq}
\end{figure}
\clearpage

\begin{figure}[p]
    \centering
    \includegraphics[width=\textwidth]{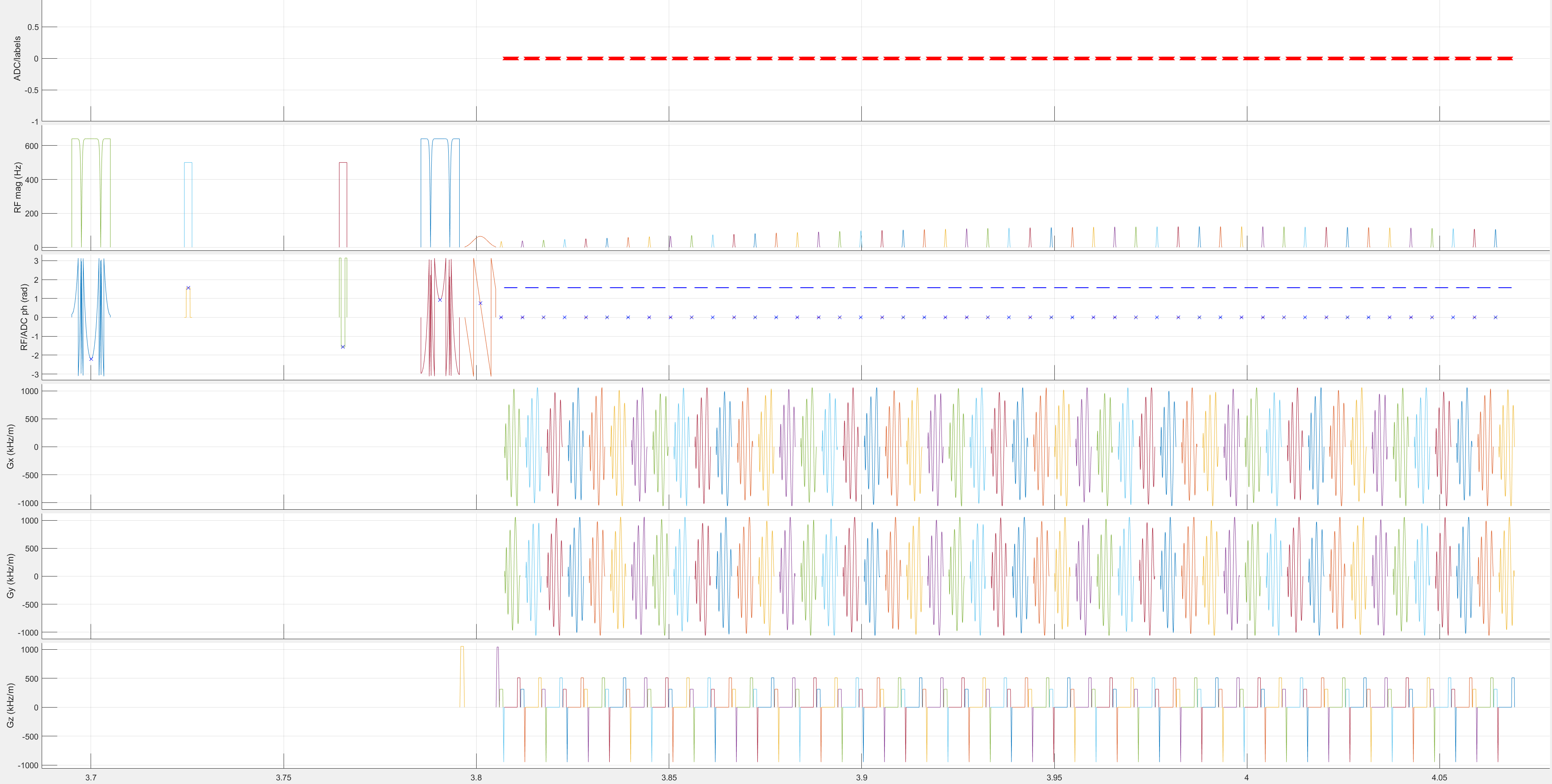}
    \caption{Sequence diagram of the $T_2$ preparation module used in the cMRF sequences. The preparation utilizes adiabatic excitation (BIR-4) and composite refocusing pulses for improved B0 and $B_1^+$ robustness.}
    \label{sfig:t2prep}
\end{figure}
\clearpage

\begin{figure}[p]
    \centering
    \includegraphics[width=\textwidth]{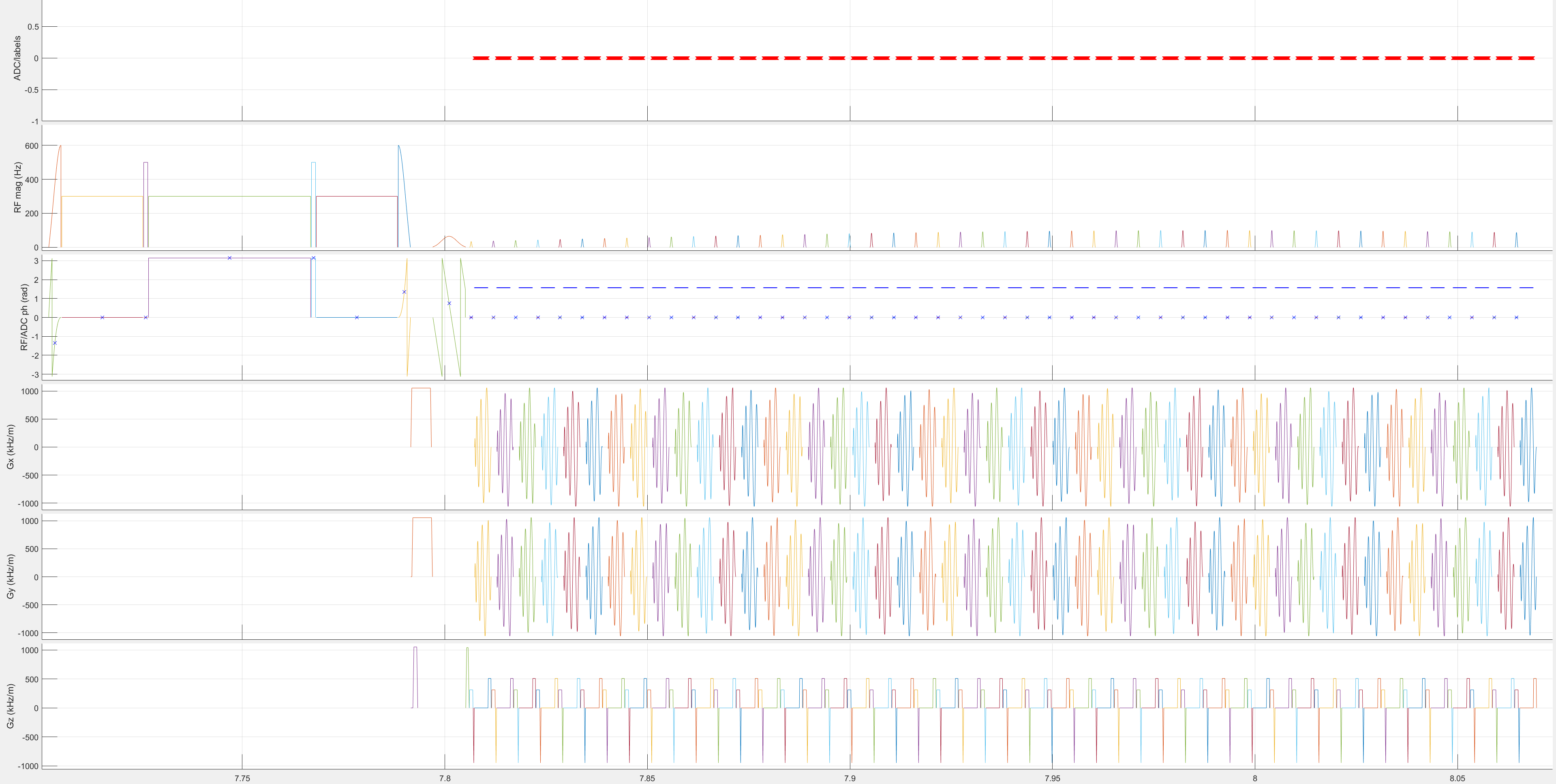}
    \caption{Sequence diagram of the $T_{1\rho}$ preparation module used in the cMRF sequences. The preparation utilizes adiabatic half passage excitation pulses and rotary-echo phase cycling based on the balanced spin-lock concept for improved B0 and $B_1^+$ robustness.}
    \label{sfig:t1rho-prep}
\end{figure}
\clearpage

\begin{figure}[p]
    \centering
    \includegraphics[width=\textwidth]{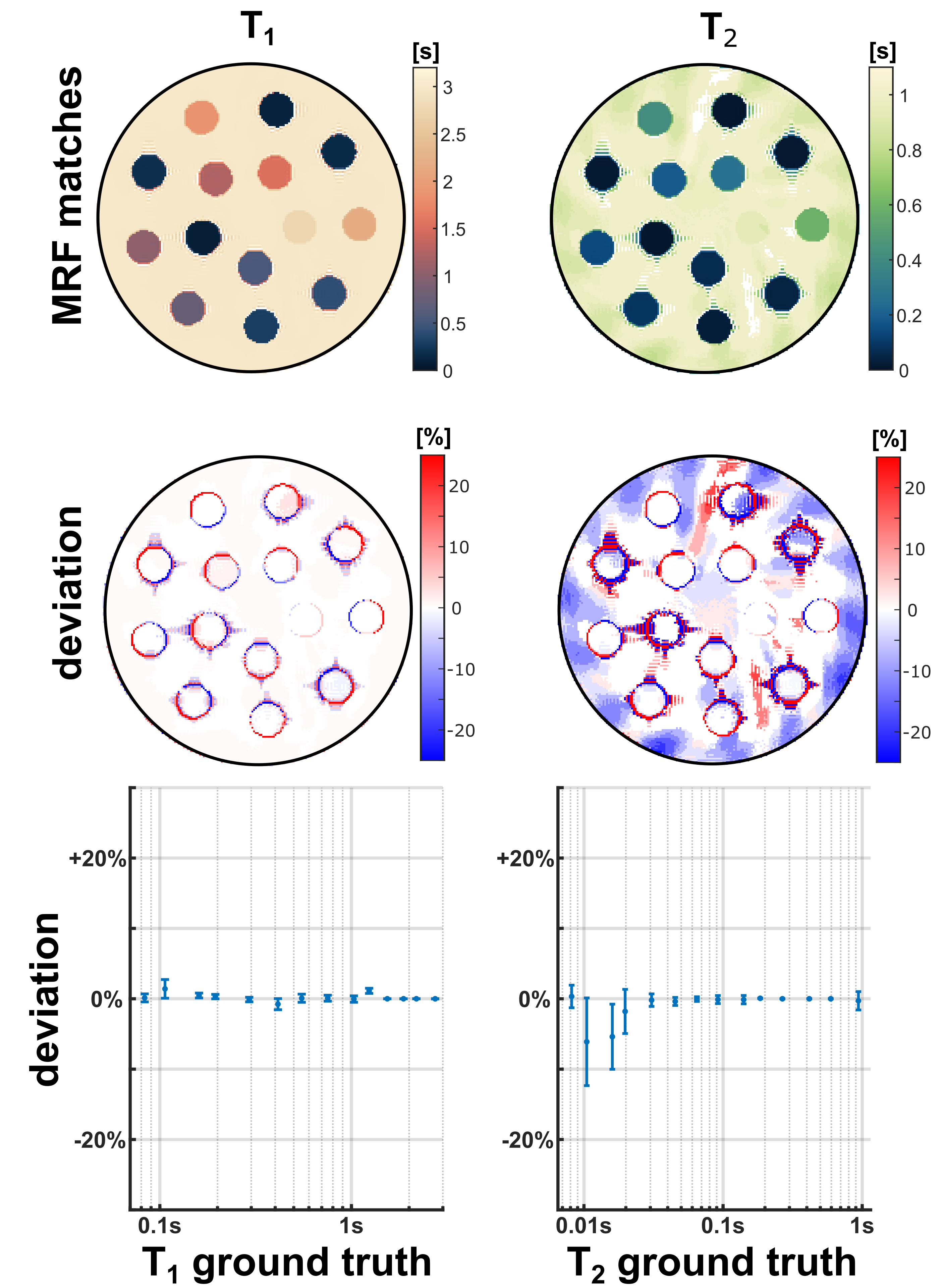}
    \caption{Digital phantom simulation of the $T_1$-$T_2$-cMRF sequence. The reconstruction was carried out using the low-rank approach. The estimated parameters are shown in the upper row. Deviations from the ground truth are depicted in the middle row and the deviations evaluated inside the individual inserts are shown in the bottom row. $T_2$ shows increased deviations in the phantom background and in inserts with short and long relaxation times. The averaged deviations are $0.19 \pm 0.74$ \% ($T_1$) and $-1.08 \pm 3.07$ \% ($T_2$).}
    \label{sfig:digital-tt-cmrf}
\end{figure}
\clearpage

\begin{figure}[p]
    \centering
    \includegraphics[width=\textwidth]{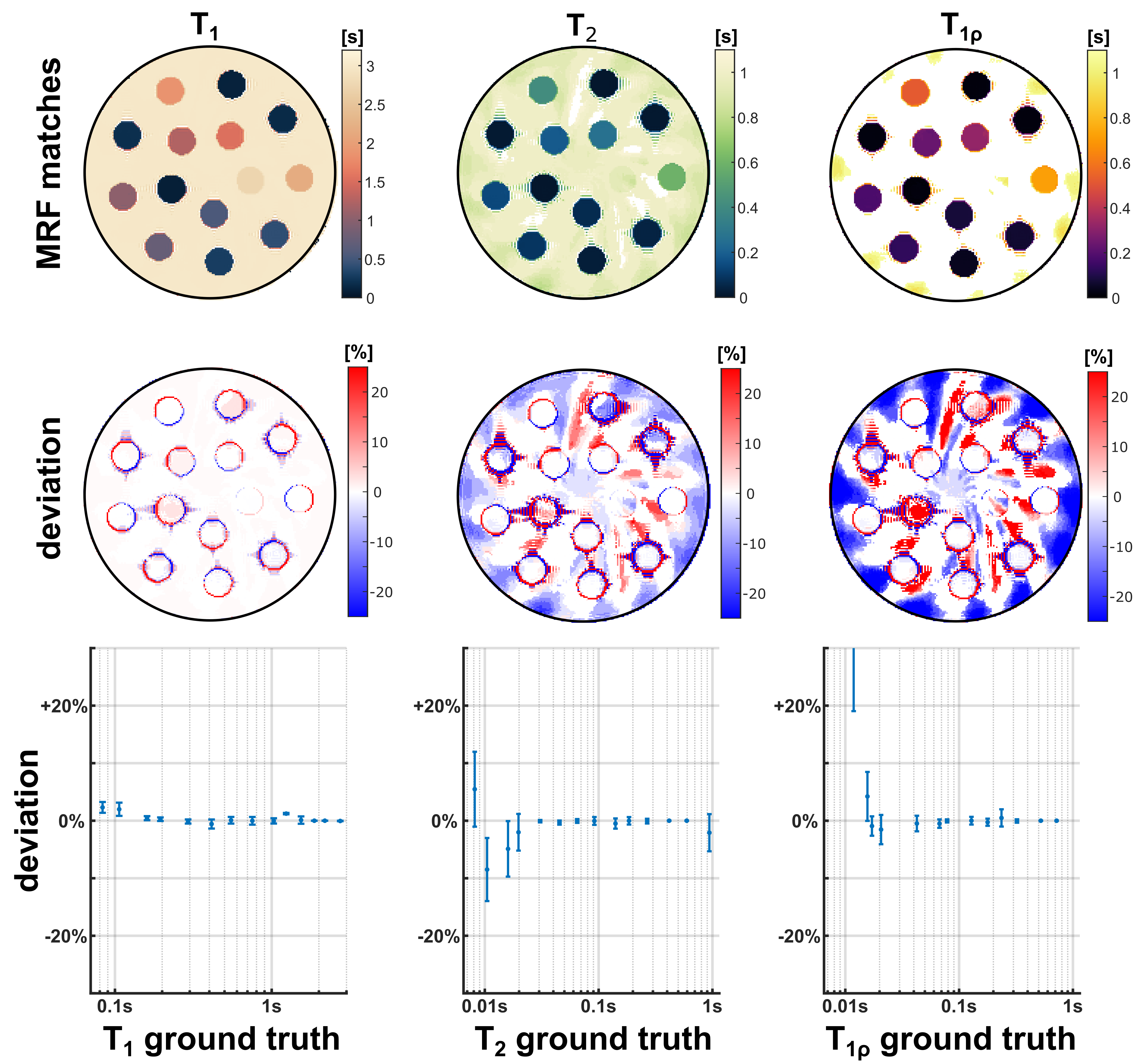}
    \caption{Digital phantom simulation of the $T_1$-$T_2$-$T_{1\rho}$-cMRF sequence. The reconstruction was carried out using the low-rank approach. The estimated parameters are shown in the upper row. Deviations from the ground truth are depicted in the middle row and the deviations evaluated inside the individual inserts are shown in the bottom row. $T_2$ and $T_{1\rho}$ show increased deviations in the phantom background and in inserts with short and long relaxation times. The averaged deviations are $0.40 \pm 0.99$ \% ($T_1$), $-0.94 \pm 4.15$ \% ($T_2$) and $-0.10 \pm 0.84$\% ($T_{1\rho}$).}
    \label{sfig:digital-ttrho-cmrf}
\end{figure}
\clearpage

\begin{figure}[p]
    \centering
    \includegraphics[width=\textwidth]{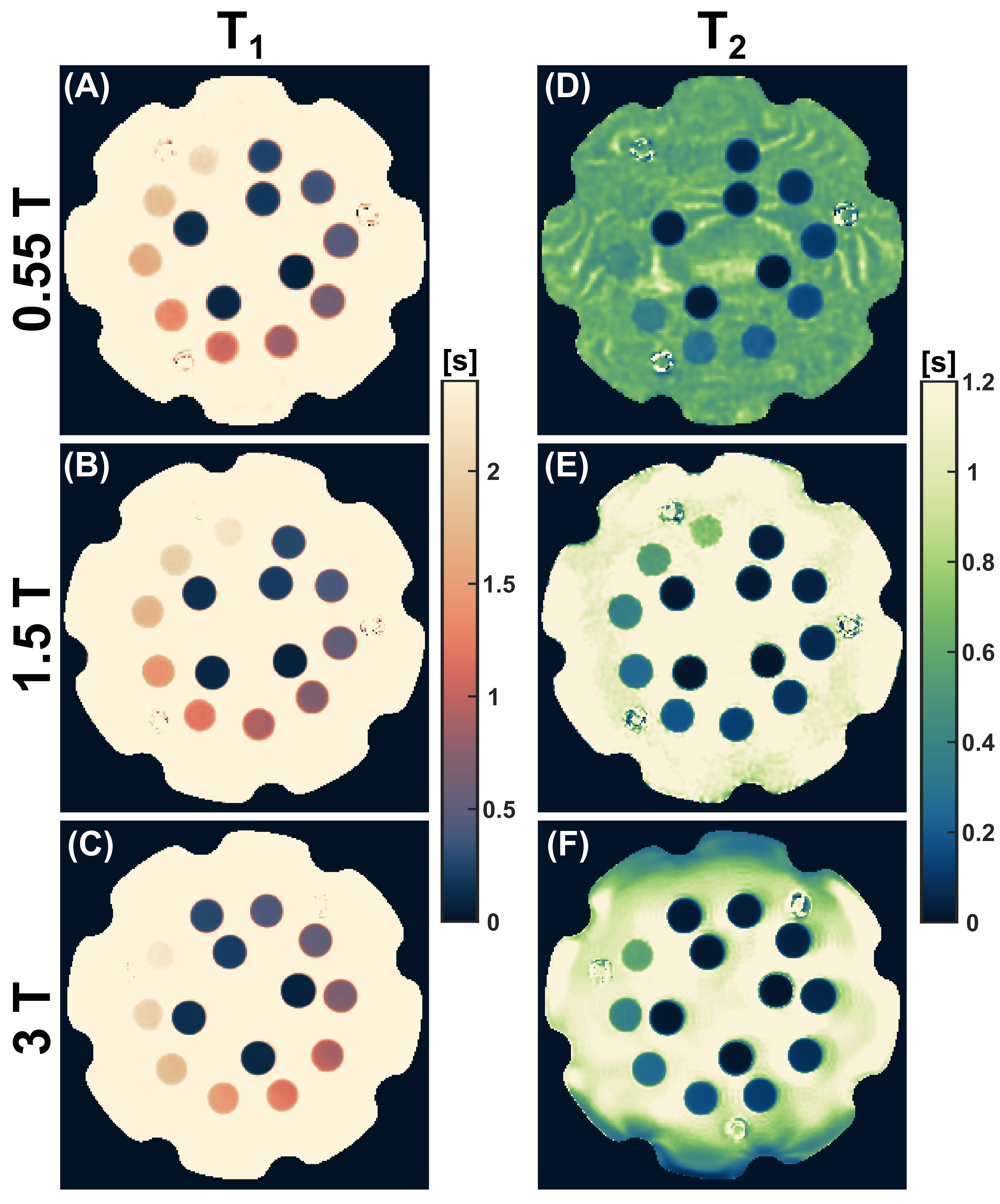}
    \caption{Multi-field-strength validation of the $T_1$-$T_2$-$T_{1\rho}$ cardiac MRF (cMRF) sequence in the ISMRM/NIST system phantom, acquired using Siemens Free.Max (0.55\,T), Sola (1.5\,T), and Vida (3\,T) scanners. All datasets were reconstructed using the low-rank subspace method. The dictionary was generated with $T_1$ values ranging from 10\,ms to 4\,s in 5\% steps, and $T_2$ and $T_{1\rho}$ values ranging from 1\,ms to 3\,s in 5\% steps. Only physically relevant combinations were used ($T_2$ $<$ $T_1$, $T_{1\rho}$ $<$ $T_1$, and $T_2$ $<$ $T_{1\rho}$). A--C) Reconstructed $T_1$ maps, D--F) $T_2$ maps, and G--I) $T_{1\rho}$ maps at 0.55\,T, 1.5\,T, and 3\,T, respectively. Simultaneous estimation of three relaxation parameters ($T_1$, $T_2$, $T_{1\rho}$) was successfully achieved. $T_{1\rho}$ values are slightly higher than $T_2$ due to small dispersion effects. Image quality is generally high across field strengths, though increased noise is visible in the $T_{1\rho}$ map at 0.55\,T and slight banding artifacts appear in the $T_2$ map at 3\,T.}
    \label{sfig:b1-wasabi}
\end{figure}
\clearpage

\begin{figure}[p]
    \centering
    \includegraphics[width=\textwidth]{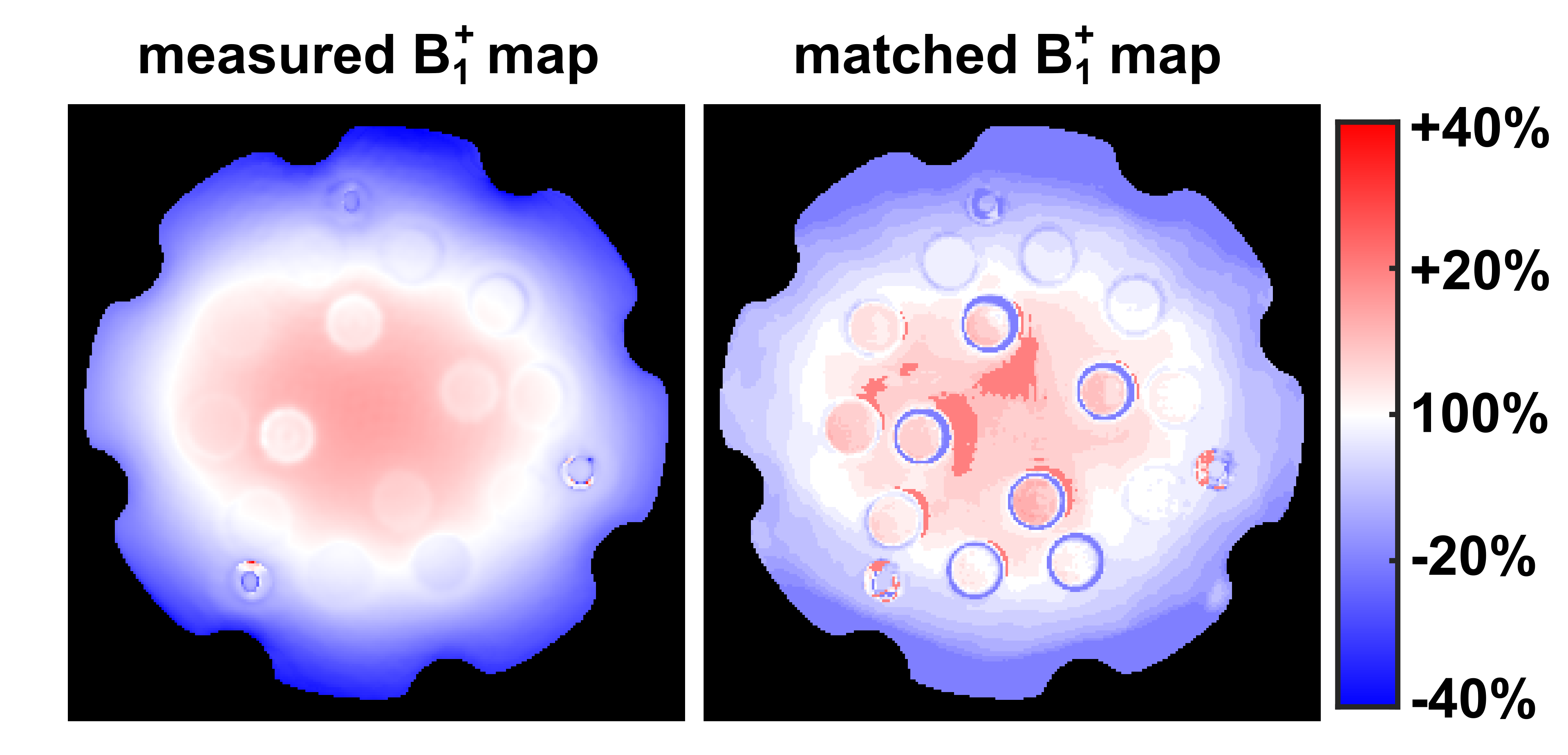}
    \caption{Comparison of a reference $B_1^+$ map obtained with the WASABI method and a matched $B_1^+$ map estimated using the $T_1$-$T_2$-cMRF sequence. Data were acquired in the ISMRM/NIST phantom on a 3\,T Siemens Prisma scanner. The MRF-derived $B_1^+$ map closely reproduces the spatial structure of the reference, capturing the characteristic central brightening associated with increased RF transmission in the phantom center. The $B_1^+$ effect in the MRF dictionary was modeled on a coarse grid with 2.5\% increments. Notably, the WASABI reference exhibits reduced $B_1^+$ values in inserts with short relaxation times, caused by relaxation during the WASABI preparation pulse, which was not accounted for in the fit routine. For these inserts, the MRF-estimated $B_1^+$ values can therefore considered more accurate.}
    \label{sfig:b1-match}
\end{figure}
\clearpage

\begin{figure}[p]
    \centering
    \includegraphics[width=\textwidth]{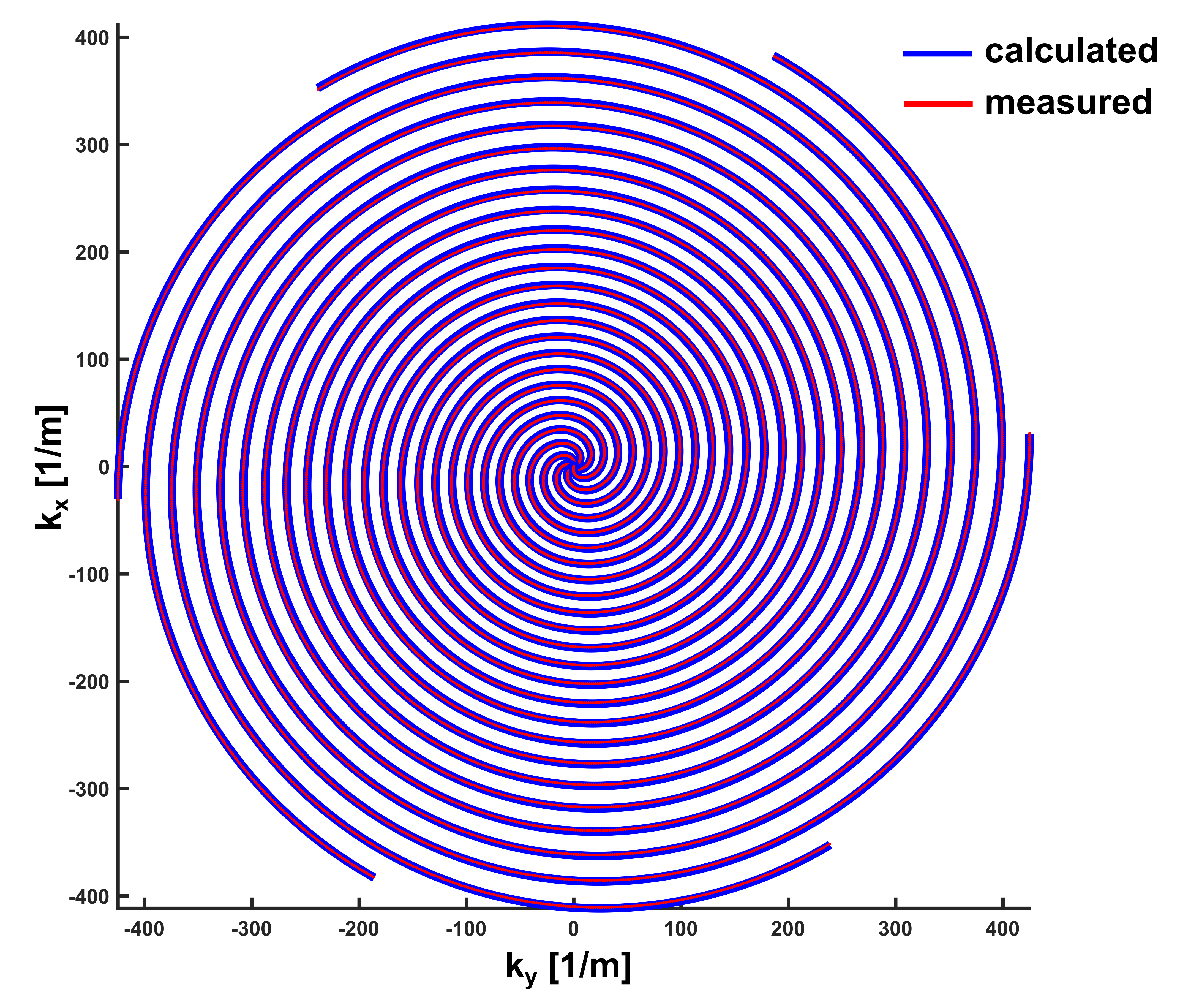}
    \caption{Comparison of the calculated and measured k-space trajectory for the IR-FISP MRF sequence for 6 spiral interleaves. The calculation of the k-space trajectory was done using Pulseq’s default tools. The measurement was carried out using the approach described in \cite{Robison2019} in a spherical calibration phantom. The .seq file for this calibration step can be generated automatically using OpenMRF.}
    \label{sfig:traj-comparison}
\end{figure}
\clearpage

\begin{figure}[p]
    \centering
    \includegraphics[width=\textwidth]{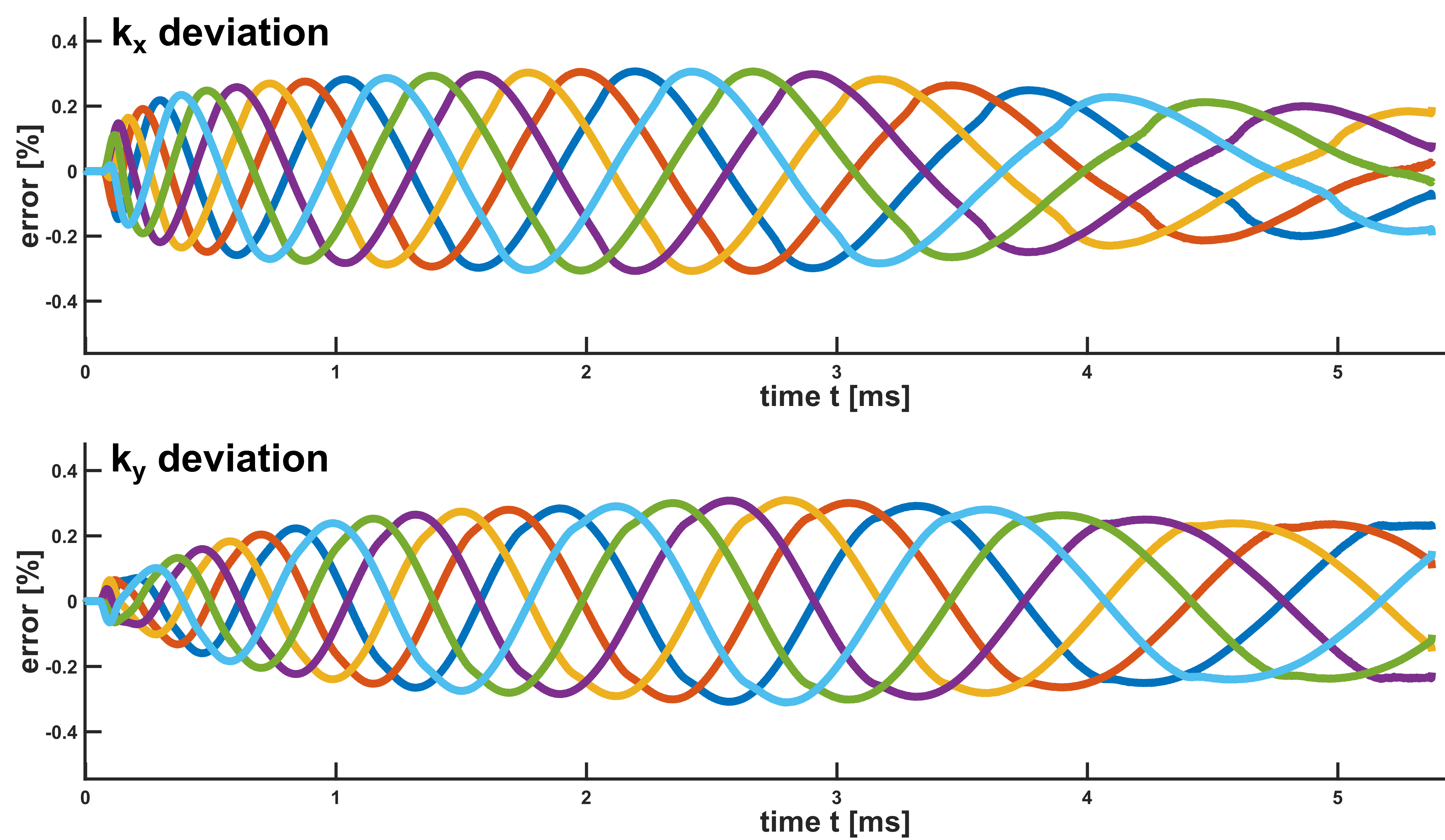}
    \caption{Deviation of the measured k-space trajectory from the calculated reference. The deviation is shown as the relative error compared to the nominal $k_{\max}$ of the entire trajectory (426.67\,1/m).}
    \label{sfig:traj-deviation}
\end{figure}
\clearpage

\begin{figure}[p]
    \centering
    \includegraphics[width=\textwidth]{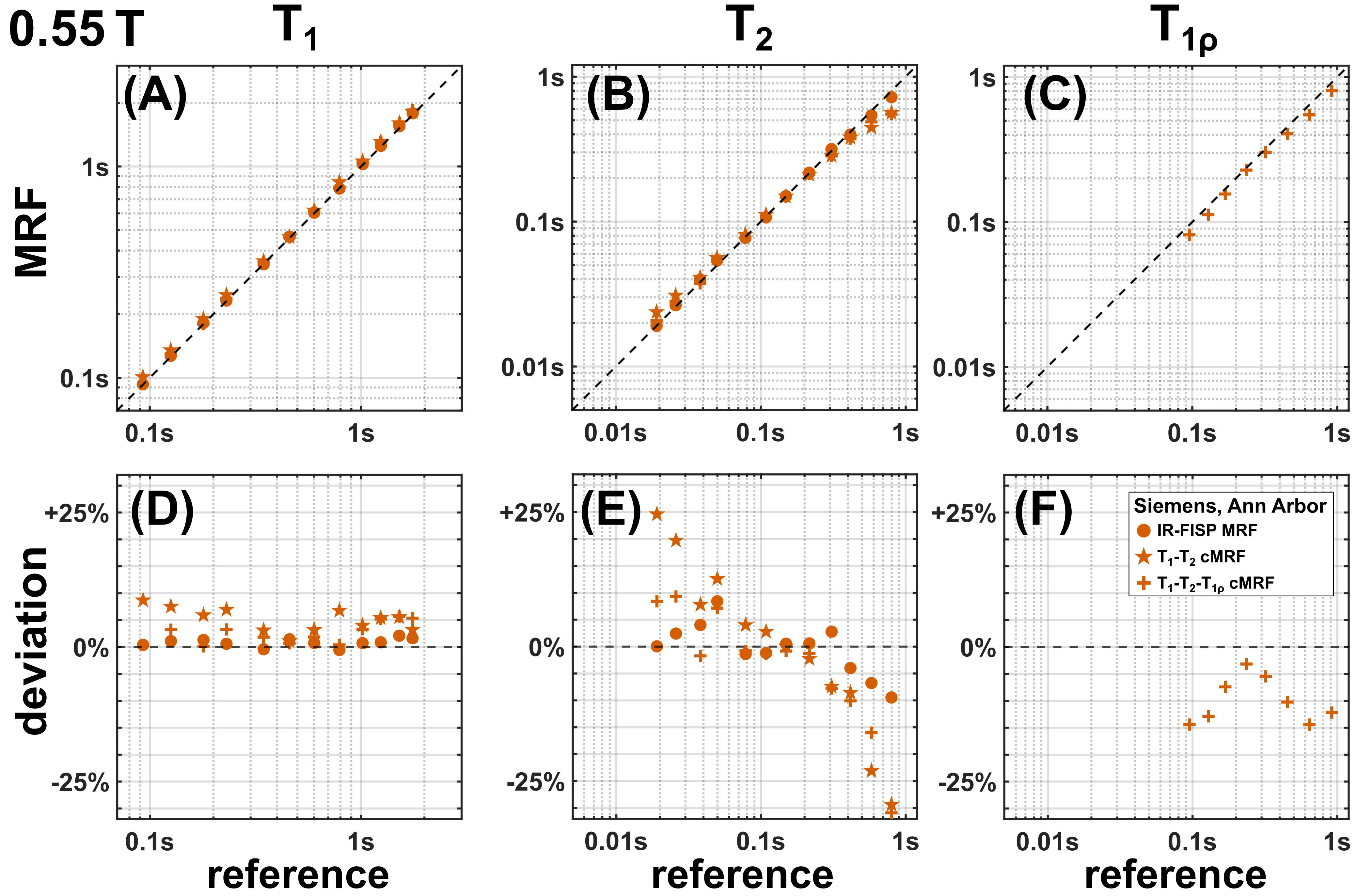}
    \caption{Quantitative comparison of $T_1$, $T_2$, and $T_{1\rho}$ estimates obtained with three MRF sequences (IR-FISP MRF, $T_1$-$T_2$ cMRF, and $T_1$-$T_2$-$T_{1\rho}$ cMRF) in the ISMRM/NIST system phantom at 0.55\,T. All measurements were acquired on a Siemens Free.Max scanner at the University of Michigan. A--C) Estimated relaxation times plotted against the gold-standard reference values for $T_1$, $T_2$, and $T_{1\rho}$, respectively. D--F) Corresponding percentage deviations relative to the reference measurements. $T_1$ estimates remain within a small deviation range across sequences, $T_2$ estimates exhibit larger variability, and $T_{1\rho}$ estimates show a slight bias toward smaller values.}
    \label{sfig:phantom-055t}
\end{figure}
\clearpage

\begin{figure}[p]
    \centering
    \includegraphics[width=\textwidth]{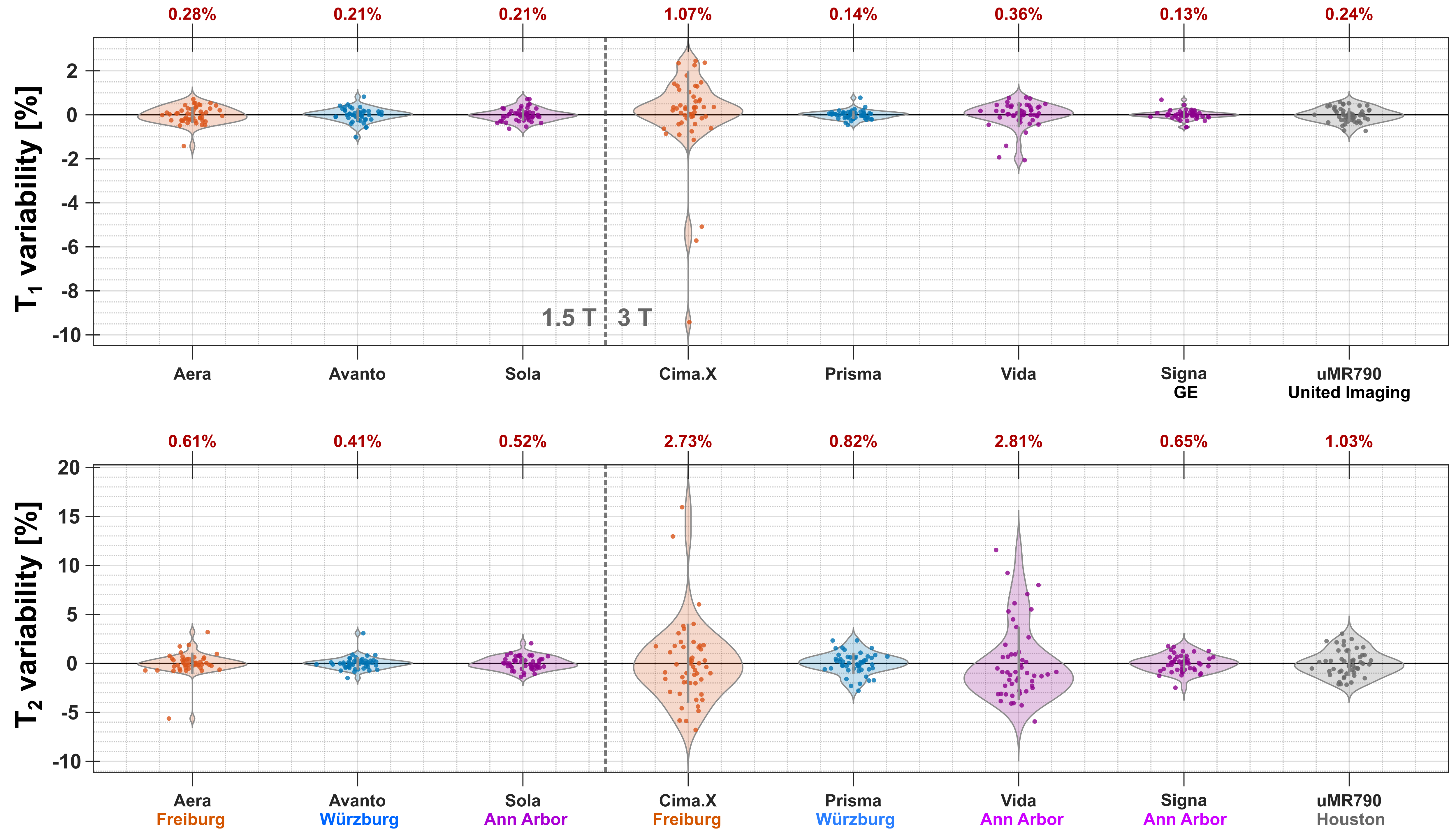}
    \caption{Violin plots with overlaid scatter points show the distribution of relative deviations for $T_1$ (top) and $T_2$ (bottom) measurements across eight MRI platforms at 1.5\,T and 3\,T. Data were acquired using IR-FISP MRF sequence. For each platform, five repeated measurements were performed on the ISMRM/NIST system phantom, and regions of interest corresponding to spheres \#3--12 were evaluated. For each sphere and platform, a reference value was defined as the mean across the five repetitions. Relative deviations were then calculated for each measurement as the percentage difference from this platform-specific reference. All deviations were pooled across repetitions and spheres to form the distributions shown for each platform. The reported variability (red labels) corresponds to the mean absolute deviation. Platforms are grouped by field strength (1.5\,T and 3\,T, separated by a dashed line) and color-coded by acquisition site. The global mean variability for $T_1$ was 0.23\% at 1.5\,T and 0.39\% at 3\,T, and for $T_2$ 0.51\% at 1.5\,T and 1.61\% at 3\,T. These results demonstrate high repeatability of the IR-FISP MRF sequence across different vendors and field strengths, with increased variability observed for selected platforms, particularly for $T_2$ at 3\,T.}
    \label{sfig:repeatability}
\end{figure}
\clearpage

\end{document}